\documentclass[reprint
              , superscriptaddress
              , aps
              , pra
              , twocolumn
              ]{revtex4-2}

\usepackage{hyphenat}

\usepackage{soul}

\usepackage[utf8]{inputenc}
\usepackage{amsmath, amsfonts, amssymb}
\usepackage[english]{babel}

\usepackage{graphicx}  
\usepackage{subfigure}
\usepackage{dcolumn}   
\usepackage{bm}        
\usepackage{multirow}
\usepackage{mathrsfs}
\usepackage{xcolor}
\usepackage{braket}
\usepackage{array}
\usepackage{rotating}
\usepackage[nolist]{acronym}

\usepackage{natbib}
\setcitestyle{numbers}
\setcitestyle{square}

\newcommand{\tj}[6]{
	\begin{pmatrix}
		#1 & #2 & #3 \\
		#4 & #5 & #6
	\end{pmatrix}
}

\begin{document}

\title{Calculation of electron-impact and photo-ionization cross sections of methane using analytical Gaussian integrals}

\author{Abdallah Ammar}
\affiliation{Universit\'e de Lorraine, CNRS, Laboratoire de Physique et Chimie Théoriques, F-57000 Metz, France}
\affiliation{Present postal address: Universit\'e de Toulouse, CNRS, Laboratoire de Chimie et Physique Quantiques, 31062 Toulouse, France.  }

\author{Arnaud Leclerc}
\email{arnaud.leclerc@univ-lorraine.fr}
\affiliation{Universit\'e de Lorraine, CNRS, Laboratoire de Physique et Chimie Théoriques, F-57000 Metz, France}

\author{Lorenzo Ugo Ancarani}
\affiliation{Universit\'e de Lorraine, CNRS, Laboratoire de Physique et Chimie Théoriques, F-57000 Metz, France}

\begin{abstract}
The ionization by photon or electron impact  of the inner ($2a_1$) and outer ($1t_2$) valence orbitals
 of the CH$_4$ molecule is investigated theoretically.
 In spite of a number of approximations, including a monocentric approach and a rather simple distorting molecular potential, the calculated cross sections are overall similar to those of other theoretical methods, and in reasonable agreement with experimental data.
 The originality of the present approach stands in the way we evaluate the transition matrix elements. The key ingredient of the calculation scheme is that the continuum radial wave function of the ejected electron is represented by  a finite sum of complex Gaussian type orbitals. This numerically expensive optimization task is then largely compensated by rather simple and rapid analytical calculations of the necessary integrals, and thus all ionization observables, including cross section angular distributions. The proposed and implemented Gaussian approach is proved to be numerically very reliable in all considered kinematical situations with ejected electron energy up to 2.7 a.u.. The analytical formulation of the scheme is provided here for bound molecular states described by monocentric Slater type orbitals; alternatively, one may also use monocentric Gaussian type orbitals for which the formulation is even simpler. In combination with complex Gaussian functions for the continuum states, an all Gaussian approach with multicentric bound states can be envisaged.
\end{abstract}

\maketitle

\begin{acronym}
	\acro{TDCS}{Triple Differential Cross Section}
	\acro{STOs}{Slater Type Orbitals}
	\acro{GTOs}{Gaussian Type Orbitals}
	\acro{cGTOs}{complex Gaussian Type Orbitals}
	\acro{rGTOs}{real Gaussian Type Orbitals}
\end{acronym}


\section{Introduction \label{introduction}}


Atomic and molecular collision processes remain an active research area, both for their fundamental interest and for numerous applications for which data are needed.
Experimental measurements of observables provide challenges to theoreticians who attempt to solve quantum mechanical many-body problems by necessarily making some approximations.
Among the many existing processes, the photon or electron impact ionization of polyatomic molecules is a difficult one to describe theoretically (see, e.g., \cite{Tennyson_2010,madison_2010,granados_2016}).
Indeed, on top of the multielectron target structure, the multicentric nature of the problem makes well-established atomic scattering tools inadequate and inappropriate.
This is true already for small molecules such as water or methane, and is worse with molecules lacking simple symmetries and/or of increasing size.

For atomic ionization processes, experimental advances allow nowadays for measurements of angular distributions of multiply differential cross sections, thus providing stringent tests to theoretical models.
Theoretical developments have also made huge progress. For example, they allow today for an essentially exact numerical calculation of the pure three-body problem in the continuum (see, e.g., \cite{bray2012}), as with calculations of differential cross sections for the single ionization of hydrogen by electron impact (see, e.g., \cite{rescigno1999,bray2000,colgan2006}) or for the double photoionization of helium (see, e.g.,
\cite{mccurdy2004,selles2002,colgan2001,randazzo2015}).
Over the last decades, and going beyond the three-body case, the ionization of other atoms by photons or electrons has been studied through a variety of nonperturbative and perturbative methods; each of them present advantages but also limitations, necessarily entail some approximations to the many-body problem, and have different degrees of success.
Many methods exist also in the case of molecular targets.
From a theoretical point of view, one critical issue when dealing with ionization processes in molecules is to describe accurately continuum states on extended spatial domains, that is to say the wave function of an electron ejected from a complex, anisotropic, multicentric system.
Obviously, as molecular targets of increasing size are considered, the radial domain in which the interactions play a role is larger, and the evaluation of transition matrix elements may also present some challenges since integrals of oscillatory functions need to be evaluated.
Compared to the bound part of the spectrum for which very accurate quantum chemistry tools are available,
one may state that the continuum part of molecular spectra is relatively not so well mastered, except possibly for diatomic molecules.
As a result, and because of the many-body aspects of the system, calculated ionization observables are generally only in fair agreement with available experimental data. There is clearly room for improvement especially when the focus is on the angular distributions of differential cross sections which provide the most detailed information on the physical process.

The spherical symmetry of atomic targets allow for standard partial wave techniques to be implemented within scattering theory.
Such tools have to be adapted or modified in the case of molecules.
For bound states the multicentric aspects are essentially mastered,
in particular with the use of real Gaussian Type Orbitals (rGTOs) centered on different nuclei (see, e.g.,
\cite{hill2013});
the calculation of one- and two-electron integrals is greatly facilitated by a number of mathematical properties, including the Gaussian product theorem (see, e.g., \cite{shavitt1963}).
In the case of continuum states proper multicentric expansions are cumbersome to provide, and the evaluation of related integrals presents serious difficulties.
The aim of this paper is to present a method that contributes in tackling this issue.
The key idea is to represent accurately molecular continuum states with a finite sum of complex Gaussian Type Orbitals (cGTOs).
As will be demonstrated, this leads to a closed-form formulation for an efficient calculation of ionization matrix elements, at least in a one-center framework.
By extension, the mathematical properties of Gaussian functions allow us to envisage, in future developments, a similar methodology but in a multicentric approach.

Real GTOs have a widespread success in molecular bound calculations,
and a large amount of optimized sets are today available for a variety of systems.
Compared to bound states, relatively less attempts have been made to investigate and to systematically optimize rGTOs for
calculations over continuum states~\cite{kaufmann_1989,Nestmann_1990,Faure_2002,fiori_2012, mavsin_2014,marante_2014,zhu_2021,coccia_2021, wozniak_2021,witzorky_2021,gao_2021,morassut_2022}.
Because of their unsuitable functional form, they appear to be inefficient in reproducing the fast oscillatory behavior of continuum
wave function above certain energies and over sufficiently large radial domains.

In a previous work, we have investigated the advantage of using cGTOs
over the standard rGTOs in representing oscillating and non-decreasing
wave functions~\cite{Ammar_2020,ammar_phd}.
cGTOs, that is, GTOs with complex-valued exponents,
have been introduced in molecular resonances calculation in the
framework of complex basis function method~\cite{McCurdy_1978,Rescigno_1985,White_2015}.
They have been applied to photoionization of atoms and small molecules
for total cross section calculations~\cite{Yabushita_1987,Morita_2008,Yabushita_1985,Matsuzaki_2016},
and more recently, for the calculation of differential cross sections and
photoelectron angular distributions~\cite{Matsuzaki_2017_Optimization,Matsuzaki_2017_Calculation,Furuhashi_2020}.
Another field of application of cGTOs is the electron dynamics in molecular systems~\cite{kuchitsu_2006,kuchitsu_2009}.
In such case, the exponents but also the coordinates centers of the Gaussian functions are allowed to be complex values.
In~\cite{Ammar_2021_Acomplex}, we have presented a single$-$center, one$-$active$-$electron approach using cGTOs
to study molecular photoionization of water and ammonia.
The construction of cGTOs was achieved by fitting them, through a least squares technique, to a set of regular Coulomb
functions with different energies on a discrete radial grid (the optimization procedure was originally developed with rGTOs~\cite{Nestmann_1990,Faure_2002}
and recently extended to cGTOs~\cite{Ammar_2020}).
It turns out~\cite{Ammar_2021_Acomplex} that the cGTOs exponents optimized in this way can be employed for distorted wave functions in the same energy/distance ranges
without significant loss of accuracy.

In this work, we formulate, implement and apply the cGTO approach to study the simple ionization of methane (CH$_4$) by electronic or photonic
impact.
We consider the inner ($2a_1$, next highest occupied
molecular orbital (NHOMO)) and outer ($1t_2$, highest occupied
molecular orbital (HOMO)) valence orbitals.
We show that all the transition elements needed to calculate the observables in both processes can be written in closed form, and thus easily evaluated.
The focus on the methane molecule is justified in several ways.
It is the smallest hydrocarbon,
known to be the most prevalent greenhouse gas emitted on earth from human and animal activities.
Methane also presents practical interest in astrophysics, radiobiology, and in the development of technological plasma device (see, e.g., \cite{mahato2019}
 and references therein).
In this work, the interest on methane is primarily motivated by the quite vast number of fundamental
studies available in the literature
for the two processes under consideration,
see e.g. refs.
\cite{Dalgarno_1952,Chupka_1971,Backx1975,Vanderwiel1976, Marr1980,cacelli_1988,stener2002,granados_2014,granados_2016, novikovskiy2019,moitra2020}
about photoionization
and refs.
\cite{lahmam_2009,toth_2010,toth_2015, Nixon_2011,nixon_2012,lin_2014,Mir_2015,icsik_2016,chinoune_2016,aitelhadjali_2016,granados_2017,gong_2017,houamer_2017,harvey_2019,Ali_2019,ElMir_2020,xu2012}
about differential cross sections for ionization by electronic impact.
This allows for comparisons of the present cross sections results
with several experimentally available data sets
and other theoretical results to test the
 applicability and efficiency of the proposed Gaussian approach comprehensively.




The remaining of this paper is as follows.
In Sec. \ref{sec_theory} we present the theoretical framework for both photoionization and electron impact ionization. Next we describe both the initial (bound) and final (continuum) molecular states. For the latter a cGTO representation is used; a table of the optimized exponents is provided. We then proceed by giving the analytical formulae that allow one to calculate all the necessary matrix elements. Results of calculations of ionization cross sections of the inner ($2a_1$) and outer ($1t_2$) valence orbitals of CH$_4$ are presented in Sec.
\ref{sec_results}. First the proposed analytical Gaussian approach is validated with purely numerical calculations; then cross sections are compared with data sets from the literature.
Finally, a summary and some perspectives of future developments are given in Sec. \ref{sec_summary}.

Atomic units (a.u.) in which $\hbar=e=m_e=1$ are used throughout unless otherwise specified.




\section{Theoretical framework \label{sec_theory}}

In this section we present the one-active-electron framework used to study the simple ionization of
methane by electronic or photonic impact.
We first recall the standard observables of interest and then focus on the proposed Gaussian approach, leading to analytical formulae for all the integrals required in the calculations.
We will elaborate extensively on the case of electronic impact ionization while, for photoionization, only the essential formulae previously introduced in \cite{Ammar_2021_Acomplex} will be summarized here, for self-consistency.

For both photonic and electronic collisions, we consider some initial bound molecular orbital $\phi_{i}(\mathbf{r})$  (usually defined in the molecular (MOL) frame) whose ionization energy is denoted $V_i$ and occupation number ${\cal N}_i$.
The electronic structure of CH$_4$ is $(1a_1^2 2a_1^2 1t_2^6)$ and we will focus on the outer and inner valence molecular orbitals $1t_2$ and $2a_1$ for which experimental data are available.
During the collision, one electron is ejected  with  wavevector $\mathbf{k_e}$  and associated energy $E_{k_e}=k_e^2/2$.
Throughout the paper, we will use the standard partial wave expansion for the continuum wave function of the outgoing electron,
\begin{equation}
	\psi_{\mathbf{k_e}}^-(\mathbf{r}) = \sqrt{\frac{2}{\pi}}
	\sum_{l,m} \imath^l
	e^{-\imath \delta_l }
	\frac{u_{l,k_e}(r)}{k_e r} Y_l^m(\hat{r}) Y_l^{m*}(\widehat{k_e})
	\text{,}
	\label{eq:psi_f}
\end{equation}
where $u_{l,k_e}$ is the radial function, $\delta_l$ denotes the phase shift for a given angular momentum $l$
and $Y_l^m$ are the complex spherical harmonics. The notations $\hat{r}$ and $\widehat{k_e}$ stand for the solid angles associated respectively with $\mathbf{r}$ and $\mathbf{k_e}$ in the laboratory frame centered on the heaviest atom of the molecule and with the $z$ axis defined as the direction of the incident projectile.

The laboratory frame (LAB) can be rotated into the molecular frame (MOL)
by using Wigner rotation matrices $\mathscr{D}_{\tilde{m} m}^{(l)}\left( \widehat{\mathscr{R}} \right)$~\cite{edmonds1974} involving the Euler angles $\widehat{\mathscr{R}}$.
The spherical harmonics transform according to
\begin{equation}
\label{rotation}
	Y_{l}^{m}\vert_{MOL}
= \sum_{\tilde{m} = -l}^{l}
	\mathscr{D}_{\tilde{m} m}^{(l)}\left( \widehat{\mathscr{R}} \right) Y_{l}^{\tilde{m}}\vert_{LAB} \
	\text{.}
\end{equation}
Since in the considered experiments the molecular target is randomly oriented, an average over all orientations of the calculated observables must be performed before a theory-experiment comparison can be done.


\subsection{Photoionization \label{subsec:PI}}

For self-consistency, we outline the essential formulation for evaluating the two main observables
measured in molecular photoionization experiments, namely the cross section $\sigma(k_e)$ and the asymmetry parameter $\beta$.
For a detailed derivation, we refer the reader to~\cite{chandra1987,machado1990,moitra2020,Ammar_2021_Acomplex}.
During a photoionization process, an incident electromagnetic radiation with energy
$E_{\gamma}$ interacts with a molecular target,
producing an ionized molecule and an ejected electron (photoelectron) in the outgoing continuum state $\psi_{\mathbf{k_e}}^-(\mathbf{r})$:
\begin{equation*}
	\gamma + \text{CH}_4
	\rightarrow
	\text{CH}_4^+  + e^-\left( \mathbf{k_e} \right)
	\text{,}
\end{equation*}
with energy conservation
\begin{equation}
	E_{\gamma} = E_{k_e} + V_i \text{.}
	\label{eq:energyPI}
\end{equation}
The photoelectron angular distribution
is given by the differential
cross section~\cite{bethe1957,drake1996}
\begin{equation}
	\frac{d\sigma^{PI}}{d\widehat{k_e}} = {\cal N}_i
 		\frac{4 \pi^2 k_e E_{\gamma}}{c } \left| T_{i f}^{PI} \right|^2
	\text{,}
	\label{eq:PICSdiff}
\end{equation}
$c$ being the
speed of light in vacuum.
For a photon with linear polarization along the direction $\hat{\epsilon}$,
the transition elements between the initial and the final states can be written in the dipole approximation as
\begin{equation}
	T_{i f}^{PI(\mathscr{G})} =
	\braket{ \psi_{\mathbf{k_e}}^- |
	\mathcal{T}^{(\mathscr{G})}
	|\phi_{i} }
\end{equation}
where the superscript $\left( \mathscr{G} \right)$ refers to the gauge choice:
$
	\mathcal{T}^{(\mathscr{L})}
	=\hat{\epsilon} \cdot \mathbf{r}
$	
in length gauge
and
$
	\mathcal{T}^{(\mathscr{L})}
	=
	\frac{1}{\imath E_{\gamma}}
	\hat{\epsilon} \cdot \mathbf{p}
$
in velocity gauge.
%
To take into account the random molecular orientation, the common approach consists in a rotation of $\mathcal{T}^{(\mathscr{G})}$ and $\psi_{\mathbf{k_e}}^- $ into the molecular frame (see Eq.~(\ref{rotation})), followed by an average over all the possible molecular Euler angles $\widehat{\mathscr{R}}$~\cite{chandra1987,machado1990}.
For subsequent calculations, we need to write down the rotated transition operators $\mathcal{T}_{\mu}^{\left( \mathscr{G} \right) }$ in the molecular frame:
\begin{align}
	\mathcal{T}^{\left( \mathscr{L} \right) }
	&=   \sum_{\mu=-1}^{1}
	\mathscr{D}_{\mu 0}^{1} \left( \widehat{\mathscr{R}} \right)
	\underbrace{
	 \sqrt{\frac{4\pi}{3}} r \, Y_{1}^{\mu}(\hat{r})} \nonumber \\
	 &= \sum_{\mu=-1}^{1}
	\mathscr{D}_{\mu 0}^{1}\left( \widehat{\mathscr{R}} \right) \quad \mathcal{T}_{\mu}^{\left( \mathscr{L} \right) },
	\label{eq:rotated_op1}
\end{align}

and
\begin{align}
	\mathcal{T}^{\left( \mathscr{V} \right) }
	&= \sum_{\mu=-1}^{1}
	\mathscr{D}_{\mu 0}^{1}\left( \widehat{\mathscr{R}} \right)
	\underbrace{
\frac{-1}{E_{\gamma}}  	\nabla_{\mu} } \nonumber  \\
 &= \sum_{\mu=-1}^{1} \mathscr{D}_{\mu 0}^{1}\left( \widehat{\mathscr{R}} \right)   \;	 \mathcal{T}_{\mu}^{\left( \mathscr{V} \right)
}.
	\label{eq:rotated_op2}
\end{align}
In eq. \eqref{eq:rotated_op2}, $\nabla_{\mu}$ denotes the spherical tensor components of the gradient operator~\cite{edmonds1974}.
After angular averaging, the differential cross section for linearly polarized photons takes the traditional form~\cite{chandra1987,machado1990}
\begin{equation}
	\frac{d\sigma^{PI\left( \mathscr{G} \right)}}{d\widehat{k_e}}
	=
	\frac{\sigma^{\left( \mathscr{G} \right)}(k_e)}{4 \pi}
	\left[ 1 + \beta^{\left( \mathscr{G} \right)} \, P_2(\cos{(\theta)} \right]
	\text{,}
	\label{eq:asymmetry_parameter1}
\end{equation}
where the integrated cross section is
\begin{equation}
	\sigma^{\left( \mathscr{G} \right)}(k_e)
	=
	\int \frac{d\sigma^{\left( \mathscr{G} \right)}}{d\widehat{k_e}} \, d\widehat{k_e}
	=
	\sqrt{4 \pi} \, \mathcal{A}_0^{\left( \mathscr{G} \right)}
	\text{,}
	\label{eq:restotalcrosssection}
\end{equation}
and the asymmetry parameter $\beta^{\left( \mathscr{G} \right)}$ is given by
\begin{equation}
	\beta^{\left(\mathscr{G}\right)} (k_e)
	= \sqrt{5} \, \dfrac{\mathcal{A}_2^{\left( \mathscr{G} \right)} (k_e)}
		{\mathcal{A}_0^{\left( \mathscr{G} \right)} (k_e)}
	\text{.}
	\label{eq:resasymmetry}
\end{equation}
In equation \eqref{eq:asymmetry_parameter1},
$P_2(x)=\frac{1}{2}(3x^2-1)$ is the second Legendre polynomial,
and $\theta= \widehat{ (\hat{\epsilon} , {\bf k_e}) }$
is the scattering angle in the laboratory frame.
The key quantities to be computed $\mathcal{A}_{L}^{\left( \mathscr{G} \right)} (k_e)$ are given by the following sums~\cite{moitra2020,Ammar_2021_Acomplex}:
\begin{widetext}
\begin{equation}
\begin{aligned}
	\mathcal{A}_L^{\left( \mathscr{G} \right)} \left( k_e \right)
	&=
	\frac{8 \pi E_{\gamma}}{k_e \, c}
	\sum_{l_1,m_1,\mu_1} \,
	\sum_{l_2,m_2,\mu_2} \,
	(-\imath)^{l_1-l_2}  e^{\imath \left( \delta_{l_1} - \delta_{l_2} \right) }
	\mathcal{M}_{l_1,m_1,\mu_1}^{\left( \mathscr{G} \right)}
	\left( \mathcal{M}_{l_2,m_2,\mu_2}^{\left( \mathscr{G} \right)} \right)^* \\
	& \quad \times
	(-1)^{m_1-\mu_1}  \sqrt{\frac{(2l_1+1)(2l_2+1)(2L+1)}{4 \pi}}
	\tj{l_1}{l_2}{L}{0}{0}{0} \tj{1}{1}{L}{0}{0}{0}  \\
	& \quad \times
	\tj{1}{1}{L}{\mu_1}{-\mu_2}{-\mu_1+\mu_2}
	\tj{l_1}{l_2}{L}{-m_1}{m_2}{\mu_1-\mu_2} ,
\end{aligned}
\label{eq:ALke}
\end{equation}
\end{widetext}
where $\delta_l$ stand for the phase shifts of the continuum partial waves~\eqref{eq:psi_f},
$\tj{l_1}{l_2}{l_3}{m_1}{m_2}{m_3}$ is the $3j-$Wigner symbol~\cite{edmonds1974}
and $\mathcal{M}_{l,m,\mu}^{\left( \mathscr{G} \right)}$ are the gauge$-$dependent integrals
\begin{equation}
	\mathcal{M}_{l,m,\mu}^{\left( \mathscr{G} \right)} =
	\int  \left( \frac{u_{l,k_e}(r)}{r} Y_l^m(\hat{r}) \right)^*
	\mathcal{T}_{\mu}^{\left( \mathscr{G} \right) }
	\phi_i({\bf{r}}) \, d{\bf{r}} .
	\label{eq:integrals}
\end{equation}
The calculation of observables relies on an efficient evaluation of these integrals, a task that the present work aims to accomplish.
Their precise form strongly depends on the basis sets selected for the description of $u_{l,k_e}(r)$ and $\phi_i({\bf{r}})$. More insights will be given in the dedicated subsection~\ref{sec:mat_elem_calc} where it is shown that those integrals become analytical when using cGTOs expansions for the continuum radial function.


\subsection{Ionization by electronic impact \label{subsec:Ionis_elec_impac}}

We now provide the basic formulae necessary to describe the simple ionization of methane by electronic impact, \textit{i.e.}
\begin{equation*}
	e^-\left( \mathbf{k_0} \right) + \text{CH}_4
	\rightarrow
	\text{CH}_4^+ \left( \mathbf{Q} \right) + e^-\left( \mathbf{k_s} \right) + e^-\left( \mathbf{k_e} \right)
	\text{,}
\end{equation*}
where $\mathbf{k_0}$, $\mathbf{k_s}$ and $\mathbf{k_e}$ are the wavevectors of the incident, scattered
and ejected electrons, respectively.
Neglecting the recoil energy of the ionized molecule $\mathbf{Q}$, the conservation of energy is written as
\begin{equation}
	\frac{k_0^2}{2} - V_i
	                \approx \frac{k_s^2}{2} + \frac{k_e^2}{2} \text{.}
\end{equation}
The experimental data analyzed below correspond to relatively high incident and scattered energies, and to asymmetric energy sharing (\textit{i.e.}, $k_s >> k_e$). Although it entails some degree of approximation, in such kinematical conditions, we  work here in the framework of the first Born approximation and neglect exchange between the scattered and the ejected electrons.
The triple differential cross section (TDCS) is defined by~\cite{ehrhardt1986,bransden2003}
\begin{equation}
	\frac{d^3 \sigma^{(e,2e)}}{d\Omega_f d\Omega_e d E_e } =
	\frac{{\cal N}_i}{4 \pi^2} \frac{k_f k_e}{k_0}  \left| T_{i f}^{(e,2e)} \right|^2
	\text{,}
	\label{eq:TDCS_def}
\end{equation}
where the scattering amplitude is given by
\begin{equation}
	T_{i f}^{(e,2e)} = \Braket{ \Psi^f | \mathcal{V} | \Psi^i}
	\text{.}
	\label{eq:T_def}
\end{equation}
In eq. \eqref{eq:T_def},
$\Psi^i$ represents the wave function of the system (incident electron$-$neutral molecule) before ionization,
while $\Psi^f$ represents that of the system after the collision
(scattered electron$-$ejected electron$-$ionized molecule).
$\mathcal{V}$~denotes the interaction between the incident electron and the target composed of $M$ nuclei and $N$ electrons:
\begin{equation}
	\mathcal{V} = - \sum_{m=1}^{M} \frac{Z_m}{\left| \mathbf{R}_m - \mathbf{r_0} \right|}
				  + \sum_{j=1}^{N}   \frac{1}{\left| \mathbf{r}_j - \mathbf{r_0} \right|}
	\text{,}
\end{equation}
where $\mathbf{r_0}$, $\mathbf{r}_j$ and $\mathbf{R}_m$ are, respectively,
the position vectors of the incident electron, of the $j-$th bound electron
and of the $m-$th nucleus.
For a neutral target $\sum_{m=1}^{M} Z_m = N$.

In the frozen-core approximation, the neutral molecule $\text{CH}_4$
and the ionized molecule $\text{CH}_4^+$ are described by the same molecular orbitals.
This approximation leads to the single$-$active$-$electron (position $\mathbf{r} $, $N=1$) model where the dimensionality of the
integrals is reduced from $3(N+1)$ to $6$. Moreover, if we neglect the spatial distribution of the nuclei
$\mathbf{R}_m \approx \mathbf{0}$~\footnote{ If the spatial distribution of the nuclei is explicitly and properly taken into account here, \eqref{eq:T0} is modified into Eqs. (7)-(8) of  \cite{lin_2014}. In subsequent calculations, we have checked that the cross section results are not significantly modified making or not the $\mathbf{R}_m \approx \mathbf{0}$ approximation (not shown). Thus, although not rigorous, neglecting the spatial distribution of the nuclei does not affect the overall picture presented in Section II.B or the applicability of the Gaussian approach presented here. }, the scattering amplitude becomes
\begin{equation}
	T_{i f}^{(e,2e)} = \Braket{
					\psi_{\mathbf{k_e}}^- \left( \mathbf{r} \right)  \psi_s \left( \mathbf{r_0} \right)
					|  \frac{1}{\left| \mathbf{r} - \mathbf{r_0} \right|} - \frac{1}{\left| \mathbf{r_0} \right| } |
					\phi_i\left( \mathbf{r} \right) \psi_0\left( \mathbf{r_0} \right) }
	\label{eq:Tif_impac_def}
\end{equation}
where $\psi_0$, $\psi_s$ and $\psi^-_{\mathbf{k_e}}$ are the wave functions of the incident, the scattered
and the ejected electrons, respectively.
We also suppose that the incident and scattered electrons are described by plane waves.
By using Bethe’s integral, the integration over $\mathbf{r_0}$ leads to the following expression in the laboratory frame:
\begin{equation}
T_{i f}^{(e,2e)}
	=  \frac{4 \pi}{q^2}
	\Braket{\psi_{\mathbf{k_e}}^- \left( \mathbf{r} \right)
			| e^{\imath \mathbf{q} \cdot \mathbf{r}}  - 1|
			\phi_i\left( \mathbf{r} \right)}
	\text{,}
	\label{eq:T0}
\end{equation}
with the transferred momentum
\begin{equation}
	\mathbf{q} = \mathbf{k_0} - \mathbf{k_s}
	\text{.}
\end{equation}
The matrix element~\eqref{eq:T0}
depends on the orientation of the molecule
with respect to the laboratory frame defined by Euler angles
$\widehat{\mathscr{R}}$.
Since the molecules are randomly oriented in the experiments, an average over all the possible orientations has to be performed.
The explicit calculation of~\eqref{eq:T0} using Gaussian integrals will be detailed in subsection~\ref{sec:mat_elem_calc}.


\subsection{Molecular orbitals}

The molecular orbitals of methane considered in this work are expressed as linear combinations of \ac{STOs} centered on the heaviest nucleus
 following Moccia~\cite{moccia1964I}:
%
\begin{equation}
	\phi_i \left( \mathbf{r} \right)  = \sum_{j=1}^{N_i} C_{ij} \, r^{n_j-1} \,
							e^{-\zeta_j r} \, Y_{l_j}^{m_j} \left( \hat{r} \right)
	\text{.}
	\label{eq:MO_def}
\end{equation}
$Y_{l_j}^{m_j}$ denote the complex spherical harmonics where the values of $\{ l_j , m_j \}$
are chosen depending on the irreducible representation
of the tetrahedral molecular symmetry group $T_d$ and the maximum $l_j$ does not exceed $3$.
The nonlinear parameters $\{ n_j, \zeta_j \}$ were optimized together with the
geometrical equilibrium configuration by minimizing the total energy
at the Hartree$-$Fock level
where $n_j$ are restricted to integers $\leq 8$.
The linear coefficients were obtained by the usual self$-$consistent$-$field$-$procedure.
The number of \ac{STOs} actually used for molecular orbitals of methane is typically $N_i \sim 7 - 15$.
The C$-$H distance corresponding to the equilibrium configuration is 2.080 a.u.,  and the ionization energy $V_i$ is 13.71 eV and 25.05 eV for the $1t_2$ and $2a_1$ states, respectively.
%


\subsection{Continuum wave functions \label{subsection_continuum} }

The radial functions $u_{l,k_e}(r)$ in the partial wave expansion \eqref{eq:psi_f}
are the solution of the ordinary differential equation,
\begin{equation}
	\left[ -\frac{1}{2} \frac{d^2}{dr^2} + \frac{l(l+1)}{2r^2}
	+ U^{\text{mol}}(r)  \right] u_{l,k_e}(r)
	= \frac{k_e^2}{2} u_{l,k_e}(r)
	\text{,}
	\label{eq:fct_disto_DE}
\end{equation}
where $U^{\text{mol}}(r)$ is a molecular central potential felt by the ejected electron.
Although it is not a very proper approach,
the approximation of a molecular potential with radial symmetry is justified here by the tetrahedral symmetry of the CH$_4$ molecule.
A substantial advantage of using such a central potential is that the explicit calculation of the angular average over Euler angles in the $(e,2e)$ case can be bypassed~\cite{granados_phd,granados_2017}.
In this contribution, we have investigated two choices for the potential $U^{\text{mol}}(r)$.

As a very first approximation, it is possible to consider a pure Coulomb potential $U^{\text{mol}}(r) = -z/r$,
the ejected electron feeling over the whole space the asymptotic charge $z=1$.
The Coulomb phase shift is given by $\delta_l =\arg \left( \Gamma(l+1 + \imath \eta) \right)$
with the Sommerfeld parameter $\eta = -z/k_e$, and the radial functions are the
regular Coulomb functions,
\begin{widetext}
\begin{equation}
\begin{aligned}
	u_{l,k_e}(r)
	=& F_l(\eta,k_e r)  \\
	=&
	(2k_e r)^{l+1} e^{-\frac{\pi \eta}{2}}
	\frac{\left|\Gamma\left(l+1+\imath \eta \right)\right|}
	{2\Gamma\left(2l+2 \right)} e^{\imath k_e r} \,
	\mathstrut_1 F_1 \left( l+1 + \imath \eta , 2l+2 ; - 2\imath k_e r \right)
	\text{,}
	\end{aligned}
	\label{eq:RegCoulFun}
\end{equation}
where
$\mathstrut_1 F_1$ is the Kummer confluent hypergeometric function~\cite{bateman1953,gradshteyn2014}.

The Coulomb approximation is quite crude. As an improvement,  we also consider a distorted model. We assume that the ejected electron feels a distorted radial
potential, obtained as the angular average of
the static exchange potential associated with molecular orbital $\phi_i$~\cite{fernandez_2010,granados_phd,granados_2017}:
\begin{equation}
	U_i \left( \mathbf{r}, \left\{ \mathbf{R}_m \right\} \right) = -\sum_{m=1}^{M} \frac{Z_m}{\left| \mathbf{r} - \mathbf{R}_m \right|}
		+ \sum_{i'=1}^{N_{\text{MO}}} (2-\delta_{i,i'})
		\int d \mathbf{r'}
		\dfrac{ \left| \phi_{i'} \left( \mathbf{r'} \right) \right|^2}{\left| \mathbf{r} - \mathbf{r'} \right|}
	\text{.}	
	\label{eq:potential}			
\end{equation}
\end{widetext}
Index $i'$ denotes the molecular orbitals in the form \eqref{eq:MO_def} and $N_{\text{MO}}=5$
is the total number of doubly occupied orbitals for methane: $1a_1$, $2a_1$, $1t_{2x}$, $1t_{2y}$ and $1t_{2z}$.
The charge $Z_m$ is 6 for the carbon atom and 1 for each hydrogen atom while $\mathbf{R}_m$ is the position of the
$m-$th nucleus.
%
%
By applying an angular average to this anisotropic potential \cite{fernandez_2010,granados_phd}, we obtain a radial potential
$U_i^{\text{mol}}(r)$, labeled with respect to the selected ionized orbital $\phi_i$.
It turns out that using STOs as in (\ref{eq:MO_def}) for $\phi_{i'}$ such potential has an analytic expansion in terms of incomplete Gamma functions  \cite{granados_phd}.
This average model potential (see Fig. 2 in \cite{granados_2017}) is dominated by the carbon nucleus charge at small distances,
$U_i^{\text{mol}}(r) \sim -6/r$,
it possesses a local minimum at $r=2.08$ a.u. corresponding to the radial distance of the hydrogen atoms,
and it behaves as a Coulomb$-$like potential at large distances,
$U_i^{\text{mol}}(r) \sim -1/r$.
Distorted continuum wave functions $u_{l,k_e}(r)$, for a given energy range and over a finite spatial grid, can then be numerically calculated using for example the RADIAL code~\cite{salvat2019}.


\subsection{Representation of the continuum wave functions using optimal sets of complex Gaussians}

The key idea to simplify the calculations of the transition matrix elements is to employ
cGTOs to represent the radial functions, solutions of the differential equation~\eqref{eq:fct_disto_DE},
\begin{equation}
	u_{l,k_e}(r) \approx r^{l+1} \sum_{s=1}^{N}
		\left[ c_s \right]_{l,k_e}  e^{-\left[ \alpha_s \right]_{l} \, r^2}.
	\label{eq:cG_def}
\end{equation}
The exponents $\left\{ \alpha_s \right\}$ and the coefficients $\left\{ c_s \right\}$ of cGTOs are complex$-$valued;
 the real part of the exponents is positive as to ensure integrability.
Compared to the expansion proposed and used in  \cite{Ammar_2020,Ammar_2021_Acomplex},
numerical efficiency is improved by systematically introducing
the term $r^{l+1}$ that reproduces the correct behavior of $u_{l,k_e}(r)$
at small $r$ (see behavior in Eq.~(\ref{eq:RegCoulFun}) for the pure Coulomb case).
Note that the coefficients depend on the partial wave number $l$ and the wave number
$k_e$ while the exponents depend only on $l$.
In other words, for a fixed $l$, we
use a series of $N$ exponents to represent a set of radial functions $\{u_{k_e}\}_l$ with different $k_e$;
for each function  within this set, a combination of $N$ linear coefficients $\left[ c_s \right]_{l,k_e}$ is optimized
.

\subsubsection{Optimization of the exponents for Coulomb functions}

The aim is to generate suitable sets of cGTOs that incorporate the behavior of continuum wave functions required to describe
an electron ejected with an energy up to, say, $k_e^2/2 \simeq 2$ a.u..
Here we outline the numerical approach used to generate such optimal sets of complex exponents. For more technical details we refer the reader to~\cite{Ammar_2020,Ammar_2021_Acomplex}.

We consider $L+1$ sets of regular Coulomb functions~\eqref{eq:RegCoulFun}
defined as
\begin{equation}
	\mathscr{F}_{l}:\{F_{\nu}(r)=F_{l}(\eta,k_{\nu}r)\}_{\nu=1,\dots,\nu_{\max}}
	\label{eq:ensembleF}
\end{equation}
for $l=0,\dots,L$.
In each set we take $\nu_{\text{max}}=7$ regular Coulomb functions defined
on a momentum grid $k_{\nu}=0.5+0.25(\nu-1)$ a.u., $\nu=1,\dots,7$, up to a comfortable radial distance $R = 30$ a.u..
For each set $\mathscr{F}_{l}$, we optimize $N=30$ cGTOs with complex exponents $\{ \alpha_1, \dots, \alpha_N \}_l$.
The optimization is performed using a two$-$steps iterative algorithm where the exponents and the coefficients are alternatively optimized.
After picking some initial choice of exponents spanning a large interval of real parts following ref. \cite{Ammar_2020},
($i$) a linear least square optimization is applied to update the coefficients $\{c_s\}$ for the current set of exponents
and
($ii$) the set of exponents $\left\{ \alpha_s \right\}$ is updated to minimize some cost function by using a nonlinear method.
Step ($i$) is performed by a least square fitting technique, while the second step makes use of a trust region algorithm~\cite{powell2009}. The optimization of the $N=30$ complex exponents is equivalent to a $2N=60$-variable optimization in real space.
The trust region algorithm requires reasonable initial values for the Gaussian exponents.
From our experience \cite{Ammar_2021_Acomplex}, picking the initial values of the real parts so that they span a large domain (following a geometric progression, typically from $10^{-4}$ to $10^{3}$) allows the optimization to perform efficiently.
%
We iterate over steps ($i$) and ($ii$) until a reasonable convergence is reached.
The cost function was chosen as a normalized sum over the modulus of the differences between
the fitted functions \eqref{eq:ensembleF} and their cGTOs expansions \eqref{eq:cG_def}
on the radial grid.
An extra penalty function was added to this sum in order to avoid the convergence of the real part of two different exponents to the same value.
Table~\ref{tab:exponents} reports the obtained exponents for $l=0,\dots,5$, ordered according to their real parts.

Note that the convergence of the optimized exponents with respect to the number of cGTO has been checked within the selected ranges of radial distance and energy. Should the domain of interest be larger than 30 a.u., or should the electron energy reach much higher values, the basis set would necessitate using more cGTO terms \cite{Ammar_2020}.

\begin{table*}[ht]
\caption{Optimal cGTOs exponents $\{\alpha_s\}_{l}$ obtained by  fitting  the set of Coulomb functions defined
    in eq. \eqref{eq:ensembleF}, for $l=0,\dots,5$. Each column is the result of a separate optimization for fixed $l$ and momentum range $k_e \in [0.5,2]$ a.u.}
\begin{sideways}
{\footnotesize
\begin{tabular}{>\centering p{3.80cm} >\centering p{3.80cm} >\centering p{3.80cm} >\centering p{3.80cm} >\centering p{3.80cm} >{\centering\arraybackslash} p{3.80cm}}
\hline
\hline
$           \{\alpha_i\}_{l=0}    $ & $           \{\alpha_i\}_{l=1}    $ & $           \{\alpha_i\}_{l=2}    $ & $           \{\alpha_i\}_{l=3}    $ & $           \{\alpha_i\}_{l=4}    $ & $           \{\alpha_i\}_{l=5}    $ \\
\hline
$   0.00005000 -0.01418571 \imath $ & $   0.00010000 +0.01402812 \imath $ & $   0.00010000 -0.01308829 \imath $ & $    0.00074448 -0.01518077 \imath $ & $   0.00042041 +0.01801314 \imath $ & $    0.00159008 -0.01491796 \imath $ \\
$   0.00006063 +0.01413816 \imath $ & $   0.00012080 -0.01404213 \imath $ & $   0.00012560 +0.02106493 \imath $ & $    0.00093544 -0.02122741 \imath $ & $   0.00059580 -0.00566189 \imath $ & $    0.00209202 +0.01046334 \imath $ \\
$   0.00007580 -0.02104961 \imath $ & $   0.00117197 +0.02296903 \imath $ & $   0.00015773 -0.02078103 \imath $ & $    0.00118130 -0.02991378 \imath $ & $   0.00084012 -0.00382707 \imath $ & $    0.00274700 -0.01585076 \imath $ \\
$   0.00009569 -0.02614599 \imath $ & $   0.00141253 -0.02299421 \imath $ & $   0.00019674 +0.02743318 \imath $ & $    0.00146590 -0.03010560 \imath $ & $   0.00117687 +0.01814476 \imath $ & $    0.00354830 -0.02133148 \imath $ \\
$   0.00012040 +0.02151560 \imath $ & $   0.00177397 -0.02992239 \imath $ & $   0.00024550 +0.01242271 \imath $ & $    0.00179924 +0.02139539 \imath $ & $   0.00162343 +0.02219053 \imath $ & $    0.00454543 +0.01696713 \imath $ \\
$   0.00109725 +0.02830883 \imath $ & $   0.00213067 +0.03000988 \imath $ & $   0.00494628 -0.03447467 \imath $ & $    0.00214992 +0.01452533 \imath $ & $   0.00221464 -0.01683770 \imath $ & $    0.00582765 +0.02320234 \imath $ \\
$   0.00304235 -0.03408812 \imath $ & $   0.00365199 -0.03782832 \imath $ & $   0.00592351 -0.02795514 \imath $ & $    0.00444564 +0.02861971 \imath $ & $   0.00300867 +0.02838691 \imath $ & $    0.00739477 -0.03125201 \imath $ \\
$   0.00370357 +0.03574263 \imath $ & $   0.00428802 +0.03744183 \imath $ & $   0.00739901 +0.03128443 \imath $ & $    0.00604063 -0.03570812 \imath $ & $   0.00400569 -0.03754185 \imath $ & $    0.00917980 +0.02905627 \imath $ \\
$   0.00780888 -0.04616031 \imath $ & $   0.00696598 +0.04666302 \imath $ & $   0.00905837 +0.03944678 \imath $ & $    0.00718953 +0.03616597 \imath $ & $   0.00529770 -0.02184147 \imath $ & $    0.01136147 +0.02664848 \imath $ \\
$   0.00917159 -0.03746089 \imath $ & $   0.00803634 -0.04648134 \imath $ & $   0.01242687 +0.02998707 \imath $ & $    0.01111228 -0.04627314 \imath $ & $   0.00696567 -0.02880311 \imath $ & $    0.01401846 +0.03802663 \imath $ \\
$   0.01079502 +0.04311274 \imath $ & $   0.01284832 -0.06152902 \imath $ & $   0.01512439 +0.05330831 \imath $ & $    0.01300477 +0.04158879 \imath $ & $   0.00906705 -0.03734749 \imath $ & $    0.01714154 -0.03656683 \imath $ \\
$   0.01284512 +0.04536018 \imath $ & $   0.01487323 +0.06308599 \imath $ & $   0.01749815 -0.04191442 \imath $ & $    0.01882657 +0.05293167 \imath $ & $   0.01165243 +0.03330491 \imath $ & $    0.02111116 -0.03440391 \imath $ \\
$   0.01564634 -0.06386250 \imath $ & $   0.03236364 -0.09414680 \imath $ & $   0.02042890 -0.04096816 \imath $ & $    0.02307524 -0.05808865 \imath $ & $   0.01502303 +0.03781643 \imath $ & $    0.02614509 +0.04690598 \imath $ \\
$   0.02000701 +0.06416977 \imath $ & $   0.03801527 +0.09147840 \imath $ & $   0.02419689 -0.04016191 \imath $ & $    0.04462000 -0.07722711 \imath $ & $   0.01928225 -0.04347419 \imath $ & $    0.03259740 -0.03319734 \imath $ \\
$   0.03971188 -0.09612550 \imath $ & $   0.10756052 +0.10000000 \imath $ & $   0.03433416 -0.06597030 \imath $ & $    0.05215464 +0.06680243 \imath $ & $   0.02503063 +0.04484551 \imath $ & $    0.04073126 -0.04271043 \imath $ \\
$   0.04788316 +0.09422822 \imath $ & $   0.12863538 -0.10000000 \imath $ & $   0.04048638 +0.07405181 \imath $ & $    0.06215870 +0.06685741 \imath $ & $   0.03267074 -0.04880118 \imath $ & $    0.05091787 +0.05274455 \imath $ \\
$   0.11995259 -0.00066731 \imath $ & $   0.40799848 -0.02284973 \imath $ & $   0.08409949 +0.08125006 \imath $ & $    0.13001133 -0.08001647 \imath $ & $   0.04239960 +0.05076988 \imath $ & $    0.06407541 -0.03829643 \imath $ \\
$   0.31219044 -0.01179928 \imath $ & $   0.80630966 -0.01540240 \imath $ & $   0.10076304 -0.09006168 \imath $ & $    0.19792406 +0.03912017 \imath $ & $   0.05504525 -0.05274282 \imath $ & $    0.08101939 +0.02631896 \imath $ \\
$   0.41248972 -0.02977216 \imath $ & $   1.42927640 +0.02829707 \imath $ & $   0.25398064 +0.01611193 \imath $ & $    3.48304664 +0.05637726 \imath $ & $   0.07237293 +0.04809387 \imath $ & $    0.12525854 -0.00292654 \imath $ \\
$   0.65781927 -0.01288056 \imath $ & $   2.42327576 -0.01308467 \imath $ & $   2.32580738 +0.00801313 \imath $ & $    6.86404558 -0.00065871 \imath $ & $   0.09492797 -0.03009527 \imath $ & $    1.70983371 +0.00190982 \imath $ \\
$   1.10405723 +0.01054021 \imath $ & $   4.17989941 -0.00529497 \imath $ & $   4.12895100 +0.02055836 \imath $ & $    13.2671809 +0.05008567 \imath $ & $   0.12551795 +0.00594159 \imath $ & $    3.27872310 -0.02722415 \imath $ \\
$   1.82448445 -0.03362815 \imath $ & $   7.09069517 +0.01188706 \imath $ & $   7.08198060 +0.03679987 \imath $ & $    25.6292404 -0.00211271 \imath $ & $   0.33145165 -0.01122423 \imath $ & $    6.18219569 +0.01139916 \imath $ \\
$   2.99582036 +0.02045485 \imath $ & $   12.0806570 -0.01456865 \imath $ & $   12.0813162 -0.01836618 \imath $ & $    49.5146115 -0.02630311 \imath $ & $   2.54850868 -0.03077083 \imath $ & $    11.7102351 +0.00698450 \imath $ \\
$   4.96305989 -0.02483028 \imath $ & $   20.5501915 -0.02390268 \imath $ & $   20.5708178 +0.01591283 \imath $ & $    95.8549284 +0.04348388 \imath $ & $   5.52583688 -0.01131074 \imath $ & $    22.0863210 -0.00135916 \imath $ \\
$   8.18492139 +0.00685243 \imath $ & $   35.0116837 +0.05615261 \imath $ & $   34.9802907 -0.02881737 \imath $ & $    185.266166 -0.03414244 \imath $ & $   9.12577624 +0.04149285 \imath $ & $    41.7537787 -0.02067622 \imath $ \\
$   13.5484183 -0.04605671 \imath $ & $   59.5734672 -0.00589158 \imath $ & $   59.5748807 -0.00266129 \imath $ & $    358.118997 -0.01554649 \imath $ & $   14.8064101 -0.01284305 \imath $ & $    78.8080226 +0.01465835 \imath $ \\
$   22.2924232 -0.02845368 \imath $ & $   101.369615 -0.01343685 \imath $ & $   101.389639 +0.00689767 \imath $ & $    692.202357 -0.01517908 \imath $ & $   23.9819807 -0.00905058 \imath $ & $    148.723058 -0.00604370 \imath $ \\
$   36.7387542 -0.00206857 \imath $ & $   172.563707 -0.00147167 \imath $ & $   172.538803 -0.03218049 \imath $ & $    1338.10418 +0.00869231 \imath $ & $   38.5699122 -0.03765285 \imath $ & $    280.815276 +0.02658699 \imath $ \\
$   60.5948087 +0.01166132 \imath $ & $   293.739824 +0.00974641 \imath $ & $   293.764991 -0.01043223 \imath $ & $    2586.55292 -0.02161133 \imath $ & $   62.1186444 +0.04420472 \imath $ & $    529.843891 -0.01150421 \imath $ \\
$   100.027937 +0.02118414 \imath $ & $   499.979246 +0.00798314 \imath $ & $   500.002945 +0.02320950 \imath $ & $    4999.96714 -0.08398230 \imath $ & $   99.9990712 -0.07792363 \imath $ & $    1000.00347 +0.00724988 \imath $ \\
\hline 											
\hline											
\end{tabular}
\label{tab:exponents}
}
\end{sideways}	
\end{table*}

\subsubsection{Optimization of the coefficients for distorted waves}

Although the cGTOs in Table \ref{tab:exponents} are initially optimized to fit regular Coulomb functions,
we have observed that the same sets of exponents can be used to accurately reproduce distorted
radial functions in similar energy and radial ranges (and even at slightly higher energies).
Using the distorted waves arising from eq.~\eqref{eq:fct_disto_DE} for a given energy range, we need only to perform once the linear least square
optimization of the coefficients $\left\{ c_s \right\}$ (step ($i$) in the algorithm explained above), which can be done at a low computational cost.
%


\subsection{Matrix elements evaluation using closed-form integrals \label{sec:mat_elem_calc}}

Thanks to the use of~cGTOs representation (\ref{eq:cG_def}), we derive in this section closed-form expressions for all integrals required to evaluate the observables of sections~\ref{subsec:PI} and~\ref{subsec:Ionis_elec_impac}.


\subsubsection{Photoionization: dipole transition elements}

Using cGTOs allows to easily write the dipole transition elements in both
length $(\mathscr{L})$ and velocity $(\mathscr{V})$ gauges
in closed form as first demonstrated in~\cite{Ammar_2021_Acomplex,ammar_phd}.
Here we summarize the main expressions for each gauge in a more compact formalism adapted to the Gaussian fitting of the form \eqref{eq:cG_def}.
After substituting the molecular orbital~\eqref{eq:MO_def} and the partial wave expansion of the continuum wave function~\eqref{eq:psi_f},
and performing the angular integrations with standard tools \cite{edmonds1974,varshalovich1988},
the transition element~\eqref{eq:integrals} can be written as
\begin{widetext}
 \begin{equation}
	\mathcal{M}_{l,m,\mu}^{\left( \mathscr{G} \right)} =
	(-1)^{m}  c^{\left( \mathscr{G} \right)}
	\sum_{j=1}^{N_i}  C_{ij} \, \tj{l}{1}{l_j}{-m}{\mu}{m_j} \,
	\mathcal{K}_{l,l_j} \,
	\mathcal{R}_{l,l_j}^{\left( \mathscr{G} \right)} \,
	\delta_{l,l_j \pm 1} \, \delta_{m,m_j + \mu}
	\text{,}
	\label{eq:Mtransition}
\end{equation}
where the superscript indicates the
length $\mathscr{G} = \mathscr{L}$ and velocity $\mathscr{G} = \mathscr{V}$ gauges.
The different factors in \eqref{eq:Mtransition} read as follows:
\begin{equation}
	\mathcal{K}_{l,l_j}  =
	\left\lbrace
	\begin{aligned}
		&(-1)^{l_j} \sqrt{l_j}     \quad &\text{if } l=l_j-1 \text{,} \\
		&(-1)^{l_j+1} \sqrt{l_j+1} \quad &\text{if } l=l_j+1 \text{,}
	\end{aligned}
	\right.
\end{equation}
\begin{equation}
\begin{aligned}
	c^{\left( \mathscr{L} \right)}  &= -1
	\text{,} \quad
	&\mathcal{R}_{l,l_j}^{\left( \mathscr{L} \right)} &=
		\mathcal{J}_{l,k_e} \left( \zeta_j, n_j+1 \right) \, \text{,}\\
	c^{\left( \mathscr{V} \right)}  &=  \frac{1}{E_{\gamma}}
	\text{,} \quad
	&\mathcal{R}_{l,l_j}^{\left( \mathscr{V} \right)}  &=
		-\zeta_j  \mathcal{J}_{l, k_e} \left( \zeta_j, n_j \right) +
		\left( n_j-1 + b_{l \, l_j} \right) \mathcal{J}_{l, k_e} \left( \zeta_j, n_j -1 \right)	\text{,}
\end{aligned}
\label{eq:Rquantities}
\end{equation}
\end{widetext}
with
\begin{equation}
	b_{l \, l_j}  =
	\left\lbrace
	\begin{aligned}
		&-l_j \quad \text{if } l =  l_j+1 \text{,} \\
		&l_j+1 \quad \text{if } l = l_j-1 \text{.}
	\end{aligned}
	\right.
\end{equation}
The $\mathcal{R}_{l,l_j}^{\left( \mathscr{G} \right)}$ quantities in \eqref{eq:Rquantities} contain the radial integrals defined as
\begin{equation}
	\mathcal{J}_{l,k_e} \left( \zeta, n \right) =
	\int_0^{\infty} \left(u_{l,k_e}(r)\right)^*  e^{-\zeta r} r^{n} d {r}
	\text{.}
	\label{eq:RadInteg_Jdef}
\end{equation}
By substituting now the radial function $u_{l,k_e}(r)$ by its cGTOs expansion~\eqref{eq:cG_def}, these integrals are given as a sum of closed-form integrals:
\begin{equation}
\begin{aligned}
\mathcal{J}_{l,k_e} \left( \zeta, n \right)
& \approx \sum_{s=1}^{N} \left[ c_s \right]_{l,k_e}^*
	 \int_0^{\infty}  e^{-\left[ \alpha_s \right]_{l}^* \, r^2} e^{-\zeta r} r^{l+n+1} d {r}	\\
& = \sum_{s=1}^{N} \left[ c_s \right]_{l,k_e}^*
	\mathcal{G} \left( \left[ \alpha_s \right]_l^*, \zeta  , l+n+1 \right)
	\text{,}
\end{aligned}
		\label{eq:RadInteg_Jgauss}
\end{equation}
where $\mathcal{G}$ can be evaluated as
\begin{equation}
\begin{aligned}
	\mathcal{G} \left( \alpha , \gamma, n \right)
		&= \frac{ \Gamma\left(n+1\right) }{(4 \alpha)^{\frac{n+1}{2}}} \, \,
			U\left( \frac{n+1}{2} , \frac{1}{2} ; \frac{\gamma^2}{4 \alpha } \right) \\
		&= \frac{ \Gamma\left(n+1\right) }{(2 \alpha)^{\frac{n+1}{2}}} \, \,
                        e^{\frac{\gamma^2}{8 \alpha } } \,
			D_{n+\frac{1}{2}}\left( \frac{\gamma}{\sqrt{2 \alpha}} \right)
	\text{,}
	\label{eq:Gauss_U}
\end{aligned}
\end{equation}
$U(a ,b ; z)$ being the Tricomi function and $D_a(z)$
the parabolic cylinder function~\cite{bateman1953,gradshteyn2014}.
Numerical values of those two special functions can be easily computed using standard mathematical packages \cite{johansson2013}.


\subsubsection{Ionization by electronic impact}

We proceed now to the calculation of the transition matrix element~\eqref{eq:T0} in the $(e,2e)$ case.
We make use of the Rayleigh expansion in terms of the spherical Bessel function $j_{\lambda}$~\cite{edmonds1974,varshalovich1988}
\begin{equation}
	e^{\imath \mathbf{q} \cdot \mathbf{r}}  =
	4 \pi \sum_{\lambda,\mu}  \,
	\imath^{\lambda} \,  j_{\lambda}\left( qr \right) \,
	Y_{\lambda}^{\mu*}\left( \hat{q} \right)
	Y_{\lambda}^{\mu}\left( \hat{r} \right)
	\text{,}
\end{equation}
of the molecular orbital~\eqref{eq:MO_def} and of the partial wave expansion~\eqref{eq:psi_f}
with the Gaussian representation~\eqref{eq:cG_def}. The angular part of the integration is performed with standard tools~\cite{edmonds1974,varshalovich1988}.
The transition integral~\eqref{eq:T0} can then be written as the sum of two terms
\begin{equation}
	T_{i f}^{(e,2e)} = T_{i \mathbf{k_e}}^{(1)} \left( \mathbf{q} ; \widehat{\mathscr{R}} \right) -
	T_{i \mathbf{k_e}}^{(2)} \left( \mathbf{q} ; \widehat{\mathscr{R}} \right)
	\text{,}
	\label{eq:Tifdeuxtermes}
\end{equation}
where
\begin{widetext}
\begin{equation}
\begin{aligned}
	T_{i \mathbf{k_e}}^{(1)}  \left( \mathbf{q} ;
	\widehat{\mathscr{R}} \right)
	&= \frac{(4 \pi)^2}{q^2} \sqrt{\frac{2}{\pi}} \frac{1}{k_e} \sum_{j=1}^{N_i} C_{ij}
	\sum_{\nu_j=-l_j}^{l_j}
	\mathscr{D}_{\nu_j m_j}^{l_j} \left(
	\widehat{\mathscr{R}}	
	\right) \mathcal{S}_j^{\nu_j}
	\text{,}
	\label{eq:T1_def}
\end{aligned}
\end{equation}
and
\begin{equation}
\begin{aligned}
	T_{i \mathbf{k_e}}^{(2)}  \left( \mathbf{q} ;
	\widehat{\mathscr{R}}	
	\right)
	&= \frac{4 \pi}{q^2} \sqrt{\frac{2}{\pi}} \frac{1}{k_e}
	\sum_{j=1}^{N_i} C_{ij} (-\imath)^{l_j} e^{\imath \delta_{l_j}}
	\mathcal{J}_{l, k_e} \left( \zeta_j, n_j \right)
	\sum_{\nu_j=-l_j}^{l_j}
	\mathscr{D}_{\nu_j m_j}^{l_j} \left(
	\widehat{\mathscr{R}}
	\right)
	Y_{l_j}^{\nu_j}\left( \widehat{k_e} \right)
	\text{.}
	\label{eq:T2_def}
\end{aligned}
\end{equation}
Eqs. \eqref{eq:Tifdeuxtermes}, \eqref{eq:T1_def} and \eqref{eq:T2_def} take into account the rotation (\ref{rotation}) of the spherical harmonics associated with the $\hat{q} $ and the $ \widehat{k_e}$ directions into the molecular frame, with Euler angles $\widehat{\mathscr{R}}$.
In \eqref{eq:T2_def},
the radial integrals $\mathcal{J}_{l, k_e}$ are the same as those appearing in the case of photoionization, namely~\eqref{eq:RadInteg_Jdef}, and can be calculated through Eqs.~\eqref{eq:RadInteg_Jgauss} and~ \eqref{eq:Gauss_U}.
In contrast, the elements $\mathcal{S}_j^{\nu_j}$ in the first contribution~\eqref{eq:T1_def} are a little more complicated, being defined as
\begin{equation}
\begin{aligned}
	\mathcal{S}_j^{\nu_j} &=
	\sum_{l,m} e^{\imath \delta_l} Y_l^{m}\left( \widehat{k_e} \right)
	\sum_{\lambda} \imath^{\lambda-l}
	\braket{l \, m | \lambda, \, m-\nu_j| l_j \, \nu_j}
	\left( Y_{\lambda}^{m-\nu_j}\left( \hat{q} \right) \right)^*
	\mathcal{I}_{l, k_e, \lambda}\left( \zeta_j, n_j, q \right)
	\text{,}
\end{aligned}
\end{equation}
\end{widetext}
with the Gaunt coefficients denoted as~\cite{edmonds1974}
\begin{equation}
	\braket{l_1 \, m_1 | l_3 \, m_3 | l_2 \, m_2}
	\equiv \int
	Y_{l_1}^{m_1 *} \left( \hat{r} \right)
	Y_{l_3}^{m_3} \left( \hat{r} \right)
	Y_{l_2}^{m_2} \left( \hat{r} \right) \,  d \hat{r} \text{ ,}
	\label{eq:Gaunt_def}
\end{equation}
and the radial integrals defined as
\begin{equation}
	\mathcal{I}_{l,k_e,\lambda}\left( \zeta, n, q \right) =
	\int_0^{\infty} \left(u_{l,k_e}(r)\right)^*  e^{-\zeta r} \, j_{\lambda}\left( qr \right) \, r^{n} d {r}
	\text{.}
	\label{eq:RadInteg_Idef}
\end{equation}
In order to evaluate the integrals $\mathcal{I}$~\eqref{eq:RadInteg_Idef}, we use the cGTOs representation \eqref{eq:cG_def}, leading to
\begin{equation}
\begin{aligned}
	&\mathcal{I}_{l,k_e,\lambda}\left( \zeta, n, q \right)  \\
	&	\approx
	\sum_{s=1}^{N} \left[ c_s \right]_{l,k_e}^*
	\int_0^{\infty}
	e^{-\left[ \alpha_s \right]_{l}^* \, r^2}
	e^{-\zeta r} \, j_{\lambda}\left( qr \right) \, r^{n+l+1} d {r}
	\text{.}
	\label{eq:RadInteg_Idef2}
\end{aligned}
\end{equation}
The individual integrals appearing in this sum have an oscillating integrand because of the Bessel function, as investigated in \cite{Ammar_2023_integrals}.
Their evaluation may not be easy depending on the values of $\zeta, \lambda, q, n$ and $l$.
The calculation of \eqref{eq:RadInteg_Idef2} can be facilitated by applying the finite Hankel representation of the spherical Bessel functions (formulae 10.49.1 and 10.49.2 of ref. \cite{olver_2010}),
\begin{widetext}
\begin{equation}
\begin{aligned}
	j_{\lambda}(z) &= \frac{(-\imath)^{\lambda}}{2 \imath}
		\sum_{k=0}^{[\lambda/2]}
		(-1)^k  \,
		a_{2k} ( \lambda + \tfrac{1}{2} ) \,
		\left(
		\frac{ e^{\imath z}  }{z^{2k+1}} - (-1)^{\lambda} \frac{ e^{-\imath z} }{z^{2k+1}}
		\right) \\
&		+
		\frac{(-\imath)^{\lambda}}{2}
		\sum_{k=0}^{[(\lambda-1)/2]}
		(-1)^k \,
		a_{2k+1}( \lambda + \tfrac{1}{2} ) \,
		\left(
		\frac{ e^{\imath z} }{z^{2k+2}} + (-1)^{\lambda} \frac{ e^{-\imath z} }{z^{2k+2}}
		\right)
\end{aligned}
\label{eq:sphBess_comb1}
\end{equation}
where the polynomial coefficients are given by
\begin{equation}
	a_k (\lambda+\tfrac{1}{2} ) =
	\left\lbrace
	\begin{aligned}
		& \frac{(\lambda+k)!}{2^k k! (\lambda-k)!}, \quad &k&=0,1,\dots,\lambda \\
		& 0, \quad &k&=\lambda+1,\lambda+2,\dots
	\end{aligned}
\right.
\end{equation}
The square brackets in the upper summation boundaries $[x]$ denote the integer part of $x$,
and a sum is ignored if the lower summation boundary exceeds the upper one.
Substituting this expression in~\eqref{eq:RadInteg_Idef2}, we obtain
\begin{equation}
	\mathcal{I}_{l,k_e,\lambda}\left( \zeta, n, q \right) =
		\sum_{s=1}^{N} \left[ c_s \right]_{l,k_e}^*
		\left[
		\mathcal{I}_{l,k_e,\lambda}^{(1)} \left( \zeta, n, q, \left[ \alpha_s \right]_l^* \right) +
		\mathcal{I}_{l,k_e,\lambda}^{(2)} \left( \zeta, n, q, \left[ \alpha_s \right]_l^*  \right) \right]
	\text{,}
\end{equation}
where the $\mathcal{I}_{l,k_e,\lambda}^{(1,2)}$ integrals can also be expressed in terms of the $\mathcal{G}(\alpha,\gamma,n)$ integrals given by eq. \eqref{eq:Gauss_U},
%
\begin{align}
	\mathcal{I}_{l,k_e,\lambda}^{(1)} \left( \zeta, n, q, \left[ \alpha_s \right]_l^* \right)
	=& \frac{(-\imath)^{\lambda}}{2 \imath}
		\sum_{k=0}^{[\lambda/2]} \frac{(-1)^k a_{2k}\left( \lambda + \frac{1}{2} \right)}{q^{2k+1}}
		\nonumber \\
&		\times
		\left[ \mathcal{G} \left( \left[ \alpha_s \right]_l^*, \zeta - \imath q  , n+l-2k \right) - (-1)^{\lambda}
			   \mathcal{G} \left( \left[ \alpha_s \right]_l^*, \zeta + \imath q, n+l-2k \right)
		\right]
	\text{,} \\
	\mathcal{I}_{l,k_e,\lambda}^{(2)} \left( \zeta, n, q, \left[ \alpha_s \right]_l^*  \right)
	=& \frac{(-\imath)^{\lambda}}{2}
		\sum_{k=0}^{[(\lambda-1)/2]} \frac{(-1)^k a_{2k+1}\left( \lambda + \frac{1}{2} \right)}{q^{2k+2}}
		\nonumber \\
		&\times
		\left[ \mathcal{G} \left( \left[ \alpha_s \right]_l^*, \zeta - \imath q  , n+l-2k-1 \right) + (-1)^{\lambda}
			   \mathcal{G} \left( \left[ \alpha_s \right]_l^*, \zeta + \imath q, n+l-2k-1 \right)
		\right]
	\text{.}
\end{align}
\end{widetext}

Despite the heavy appearance of the above formulae and sums, each term is actually easy to evaluate because the $\mathcal{G}$ quantities just require a good routine for the parabolic cylinder function \cite{johansson2013}.
If the initial target state $\phi_i(\mathbf{r})$ is expanded over monocentric GTOs rather than STOs as in Eq. \eqref{eq:MO_def}, the radial integrations are much simpler; we refer to \cite{Ammar_2023_integrals} for the formulas and a detailed numerical investigation.

The transition amplitude obtained through \eqref{eq:Tifdeuxtermes} is then substituted in the differential cross section formula~\eqref{eq:TDCS_def}, followed in principle by Euler-angle averaging. This last step can actually be bypassed in the present calculations, thanks to the use of a  pre-averaged potential $U^{\text{mol}}(r)$ (see details in subsection \ref{subsection_continuum}). We finally get the TDCS for the electron impact ionization of methane in a very efficient way.



\section{Results \label{sec_results}}

The reliability of the proposed Gaussian approach is now being tested in realistic conditions inspired by experimental results taken from the literature. We will also compare the Gaussian results with those of other benchmark theoretical approaches using similar assumptions.
We focus on two target orbitals of methane, the highest occupied molecular orbital $1t_2$ and the next highest occupied molecular orbital $2a_1$.
%
We would like to emphasize that in the method proposed in this work the most demanding part of the calculation is the optimization of the cGTO exponents.
However, this preliminary first step has to be  performed only once for a chosen energy range. Typically, the multidimensional optimization takes one day with one CPU for each value of $l$. The subsequent steps, \textit{i.e.} the cross sections calculations, are relatively rapid thanks to the analytical formulation derived in section \ref{sec_theory}: typically,
30s of CPU time for each energy abscissa in the case of photoionization, and a few hours for one TDCS point at a given angle in the $(e,2e)$ case.


\subsection{CH$_4$ photoionization}

\begin{figure*}[]
	\includegraphics[width=\linewidth]{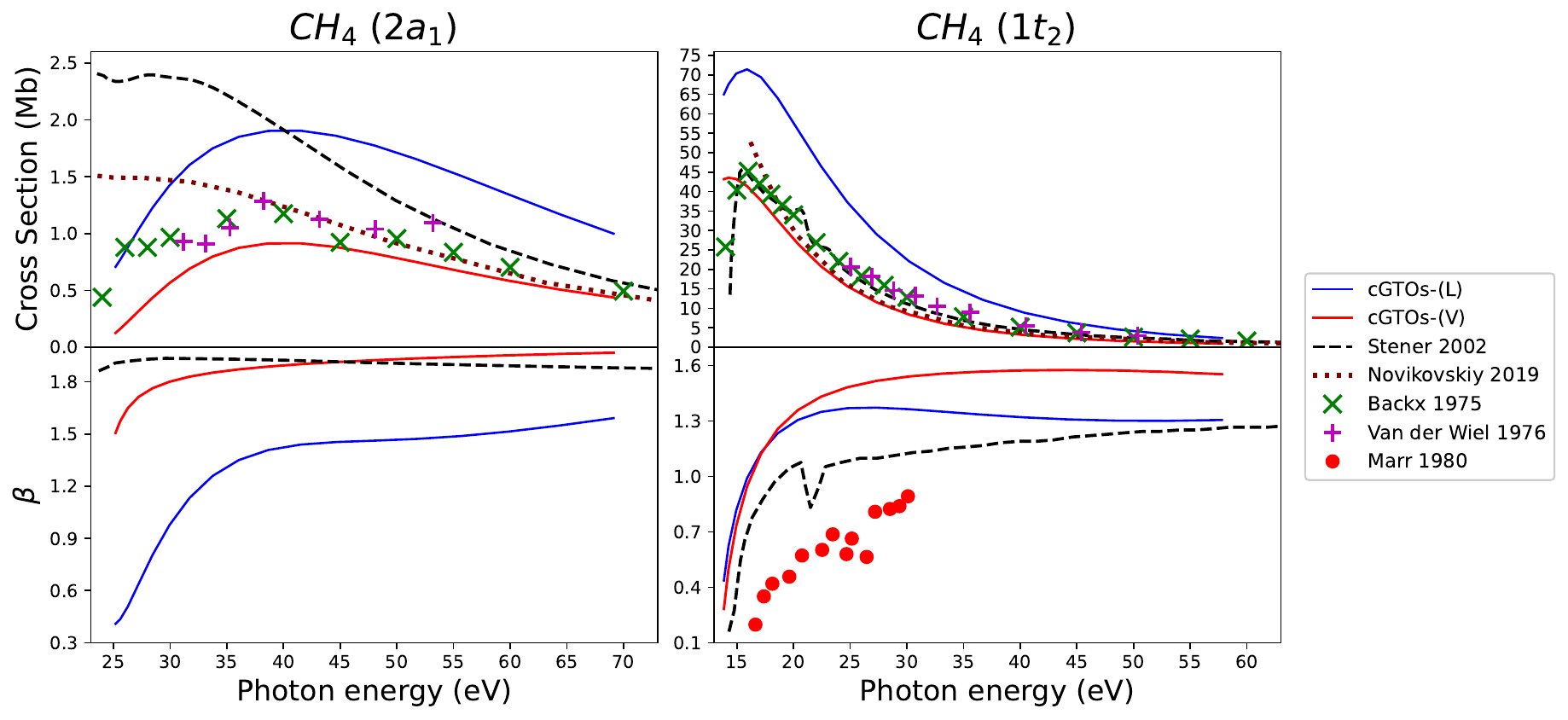}
	\caption{
	Partial photoionization cross section $\sigma^{(\mathscr{G})}(k_e)$ (top panels) and asymmetry parameter $\beta^{(\mathscr{G})}(k_e)$ (bottom panels)
	as a function of the photon energy $E_{\gamma}$ (in eV, see eq. \eqref{eq:energyPI}), for orbitals $2a_1$ (left panels) and $1t_2$ (right panels)
	of CH$_4$.
	Present results using cGTOs in both length (-L) and velocity (-V) gauges
	are compared with results
	from other theoretical methods: TD-DFT by Stener {\textit{et al}}~\cite{stener2002} (dashed line) and
	single-center method of Novikovskiy {\textit{et al}}~\cite{novikovskiy2019} (dark red dotted line),
	and experimental points:
	Backx $\&$ Van der Wiel~\cite{Backx1975} (green diagonal crosses), Van der Wiel {\textit{et al}}~\cite{Vanderwiel1976}
	(straight violet crosses) and Marr {\textit{et al}}~\cite{Marr1980} (red spots).}
	\label{fig:CH4_PI}
\end{figure*}

The photoionization cross section $\sigma^{(\mathscr{G})}(k_e)$
and asymmetry parameter $\beta^{(\mathscr{G})}(k_e)$
are presented in Fig. \ref{fig:CH4_PI}.
Results of the cGTO approach with the distorting molecular potential (see details in subsection \ref{subsection_continuum}), are shown for both length and velocity gauges,
together with available experimental points and two selected theoretical results.

For both orbitals, the Gaussian cross section obtained in velocity gauge agrees best with the experimental points (experimental uncertainties are not shown) \cite{Backx1975,Vanderwiel1976}.
The theoretical results reproduce the main features of the energy dependency: for the $2a_1$ orbital, the shape of the maximum around 40eV and the absolute values are good; for the $1t_2$ orbital, except at low photon energy, the decreasing shape and the absolute values of the results agree equally well with experimental points.

Cross sections obtained in the length gauge are larger and clearly overestimate the experimentally observed magnitude; however, the general tendencies are still correctly recovered. The Gaussian results are also compared with TD-DFT results of Stener {\textit{et al}}~\cite{stener2002} (obtained in length gauge) and
with the single-center method of Novikovskiy {\textit{et al}}~\cite{novikovskiy2019} (obtained in velocity form). None of them (including the present results) give uniformly perfect agreement but the order of magnitude of the discrepancies between theoretical with experimental points or among theoretical results remains acceptable.

The bottom panels of Fig. \ref{fig:CH4_PI} show the results for the asymmetry parameter, compared with available experimental points  of \cite{Marr1980} for the $1t_2$ orbital between 15 and 30 eV (to our knowledge, no experimental data is available outside this energy range, nor for the $2a_1$ orbital). The agreement is not so good, but this is not totally surprising because the asymmetry parameter is related to the photoelectron angular distribution which is strongly sensitive to the quality of the wave functions used in the calculation.
In contrast to what is observed for cross sections, larger absolute values are obtained in velocity gauge.
A comparison with benchmark TD-DFT results \cite{stener2002}, obtained in length gauge, shows that the general trends are similar.
Our velocity gauge results agree best with TD-DFT results in the case orbital $2a_1$, whereas length gauge results are somehow closer to the TD-DFT and to experimental points in the case of orbital $1t_2$.

A similar comparison was presented in \cite{Ammar_2021_Acomplex} for two other XH$_n$ molecules, namely water and ammonia.
In all cases, including the present calculations on CH$_4$, the cGTO description of the continuum is shown to be reliable
to calculate photoionization observables.

\subsection{CH$_4$ ionization by electronic impact in coplanar geometry}

Our investigation now turns to electronic impact ionization of methane. Several series of experimental data collected in coplanar asymmetric geometry are considered here for comparison and assessment of the cGTO calculations.

\begin{figure}[]
	\includegraphics[width=\linewidth]{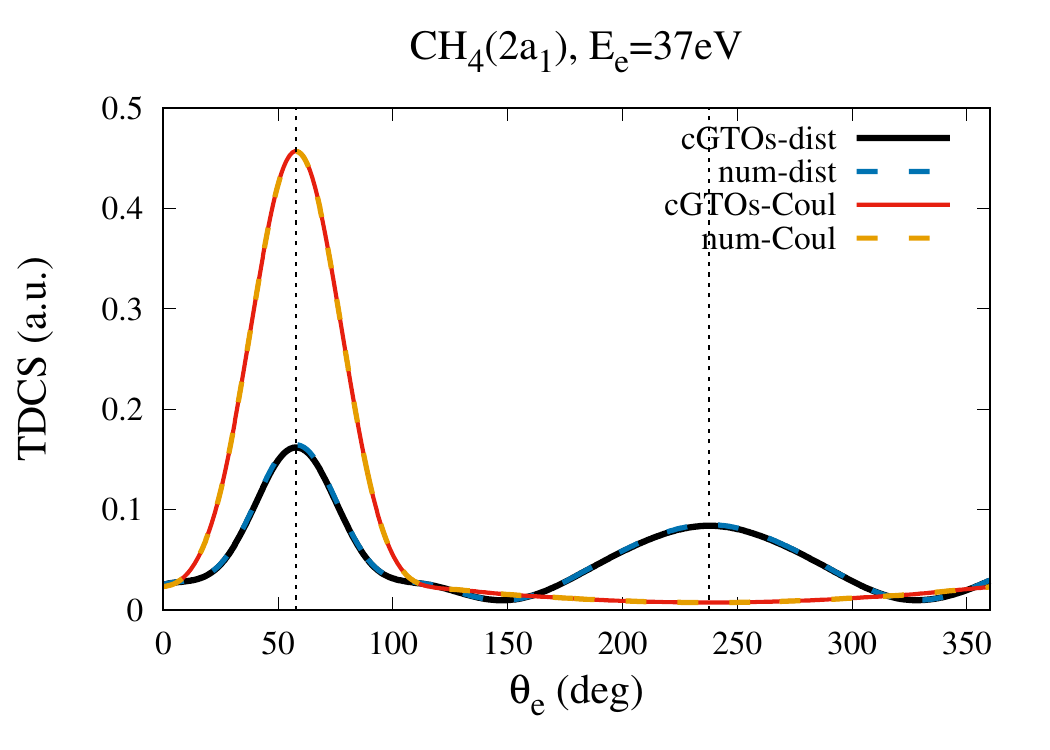}
	\caption{
\ac{TDCS} for ionization by electronic impact of the inner ($2a_1$) valence orbital of CH$_4$, as a function of ejection angle $\theta_{e}$, at fixed scattering angle of $\theta_s = -6^{\circ}$, $E_s = 500$ eV, $E_e = 37$ eV (as in the experiments of Lahmam-Bennani {\textit{et al}}~\cite{lahmam_2009}).
	For a Coulomb wave for the ejected electron (red line) or a distorted wave (black thick line), the results obtained with the cGTO approach are compared to the  purely numerical ones (blue and orange dashed lines). The vertical lines indicate the momentum transfer direction and its opposite.
		}
	\label{fig:CH4_TDCS_Lahmam_Analytical_vs Numerical}
\end{figure}

%
%
We have first calculated the TDCS for the kinematic parameters of the Lahmam-Bennani {\textit{et al}} experiment \cite{lahmam_2009} where the scattered electron is detected at an energy of 500 eV, with ejected electron energy of 12~eV, 37~eV or 74~eV, and a scattering angle of the fast outgoing electron at $\theta_s = -6 ^{\circ}$. In this experiment, the low energy electron analyzer is swept around the plane over angular ranges $\theta_{e} \in [25^{\circ},160^{\circ}] $ and $\theta_{e} \in [200^{\circ},335^{\circ}] $.
In such kinematical and geometrical configurations, the angular distributions feature two peaks: one close to the momentum transfer direction (known as binary peak) and one in the opposite direction (known as recoil peak).
%
%
The cGTO results
for the ionization of the inner $2a_1$ orbital
are shown in Fig.~\ref{fig:CH4_TDCS_Lahmam_Analytical_vs Numerical}
for an ejected electron energy $E_e = 37$ eV
with two different choices of the  continuum wave function, either a pure Coulomb wave (a quite crude choice) or a distorted wave (see details in subsection \ref{subsection_continuum}).
Calculations based on analytical integrals using cGTO expansions are compared with results obtained
with numerical integrals, that is to say using the 'exact' original wave functions and numerical quadrature. The perfect agreement shows that the cGTO representation of the continuum and the associated analytical integrals are fully reliable. The same agreement is found also for other kinematical situations and also when ionizing the 1t$_2$ orbital.
This being ascertained, all the TDCS to be presented hereafter have been obtained with the Gaussian analytical approach proposed in this work.

\begin{figure*}[]
\includegraphics[width=0.45\linewidth]{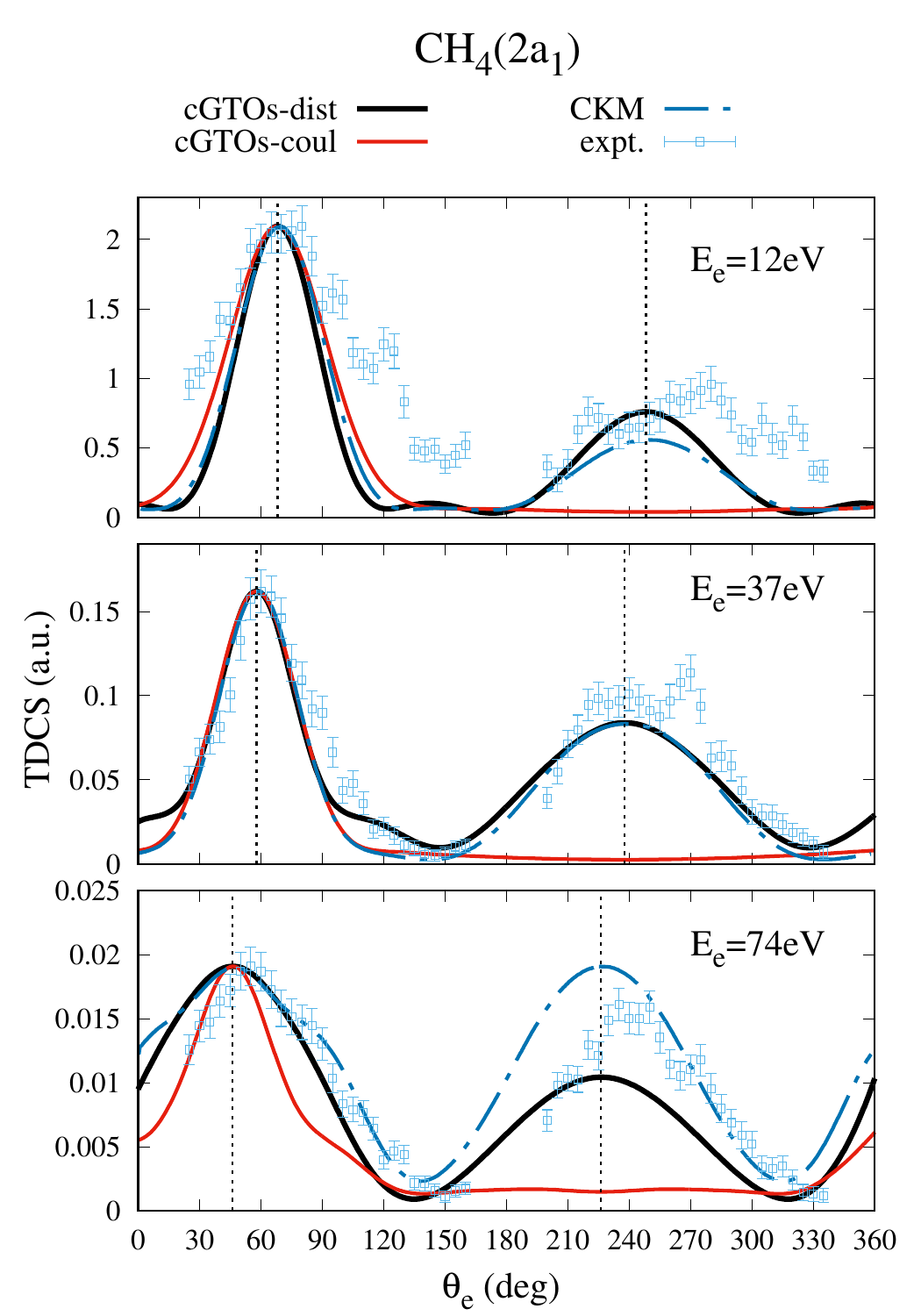}
\includegraphics[width=0.45\linewidth]{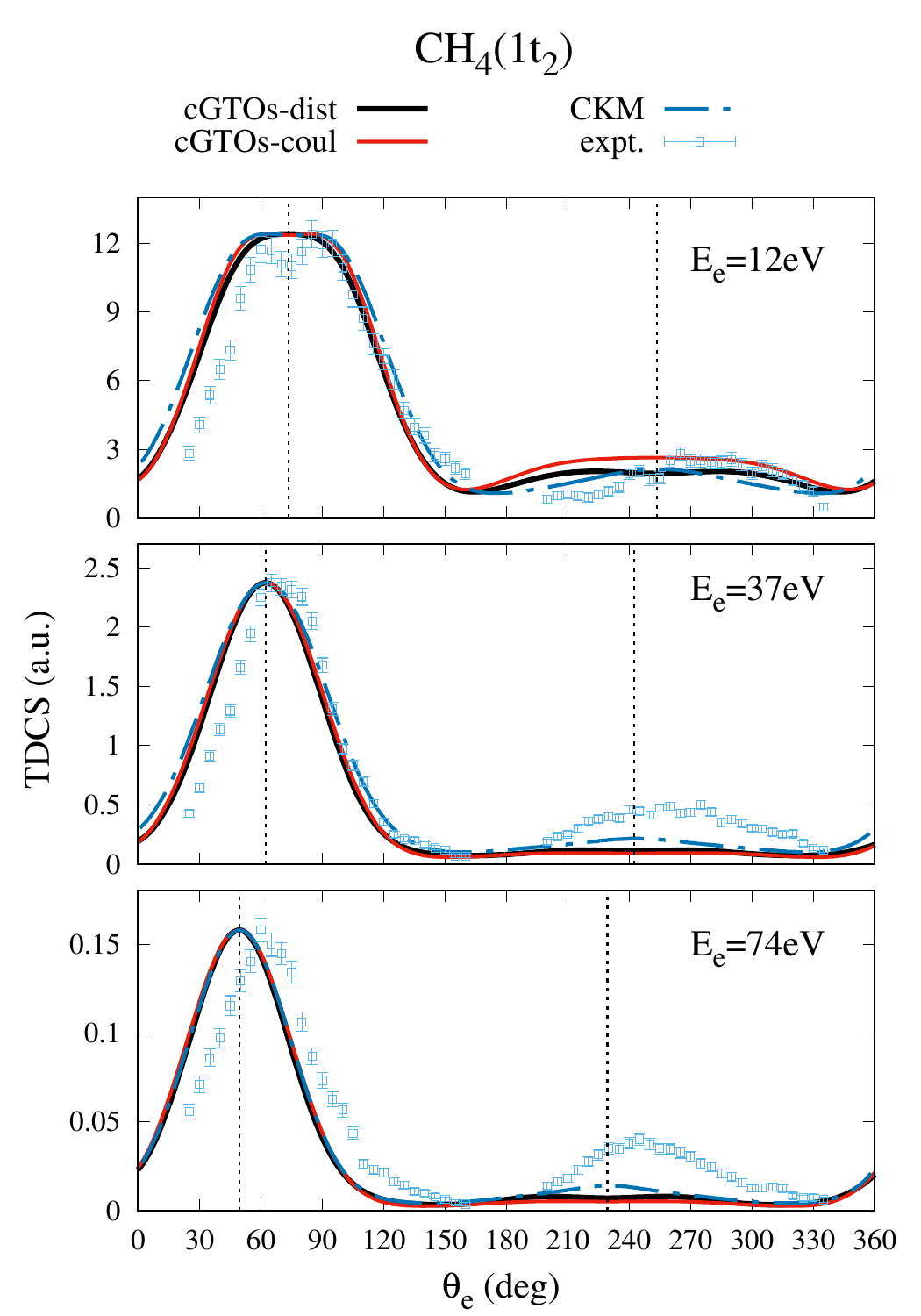}
	\caption{
\ac{TDCS} for ionization by electronic impact of the inner ($2a_1$, left panels) and outer ($1t_2$, right panels) valence orbitals of CH$_4$, as a function of ejection angle $\theta_{e}$, at fixed scattering angle of $\theta_s = -6^{\circ}$, with kinematical parameters of Lahmam-Bennani {\textit{et al}}~\cite{lahmam_2009},  $E_s = 500$ eV, $E_e = 12$ eV (upper panels), $37$ eV (middle panels) and $74$ eV (lower panels).
		The present cGTO calculations with the ejected electron described by a Coulomb wave (red line) and a distorted wave (black thick line), are compared with the complex Kohn method (CKM) results (blue dashed line) \cite{lin_2014}
and the experimental points  \cite{lahmam_2009}.
The vertical lines indicate the momentum transfer direction and its opposite. 	The relative experimental data have been normalized at the binary peak maximum to the cGTO theoretical curve obtained with the distorted potential.	}
	\label{fig:CH4_TDCS_Lahmam}
\end{figure*}

%
%

In Fig.~\ref{fig:CH4_TDCS_Lahmam} we compare theoretical and experimental results, for both the $2a_1$ and $1t_2$ orbitals, and for three values of the ejected energy.
All other theoretical curves and experimental points have been normalized to the maximum point of the distorted wave calculation at the main binary peak, independently for each panel of Fig. \ref{fig:CH4_TDCS_Lahmam}.
Our theoretical results using the Coulomb wave function
reproduce those published TDCS within the same approximation \cite{lahmam_2009,granados_2017} (not shown on Fig.~\ref{fig:CH4_TDCS_Lahmam})
\footnote{In this validation step we realized that in \cite{granados_2017}, the GSF calculations were made with an incident - instead of scattered - energy of 500 eV, that is to say a kinematical configuration slightly different from the experimental one.}.
The obtained TDCS recover the main binary peaks but clearly fail to reproduce the recoil peaks observed in the measurements.
The distorted wave calculations lead to a better agreement, especially for the $2a_1$ orbital where the relative amplitudes of the recoil and binary peaks agree quite well with the experiment.
Our distorted wave results are globally similar to those obtained within the complex Kohn method (CKM) \cite{lin_2014} in which the interaction between the ejected
electron and the residual molecular ion is treated in a close
coupling method (in \cite{lin_2014} the spatial distribution of the hydrogen nuclei is properly taken into account - see footnote [76]).
They are also very similar in both shape and magnitude (not shown) to those obtained with the multicenter distorted-wave method (MCDW)  \cite{gong_2017}  in which the continuum wave function of the slow ejected electron is calculated in
a multicenter potential of the residual ion. The similarity in calculated TDCSs indicates that the anisotropic potential plays no major role, at least for methane and for the considered geometrical and kinematical configuration.

The three calculations (present cGTO distorted wave, MCDW \cite{gong_2017} and CKM \cite{lin_2014}) provide an agreement with the measurements that is overall acceptable.
As expected from the first Born approximation, the cGTO TDCS is symmetric with respect to the momentum transfer direction and the angular shift of the experimental binary peak is not reproduced.
%
%
%

%

\begin{figure}[]
	\includegraphics[width=\linewidth]{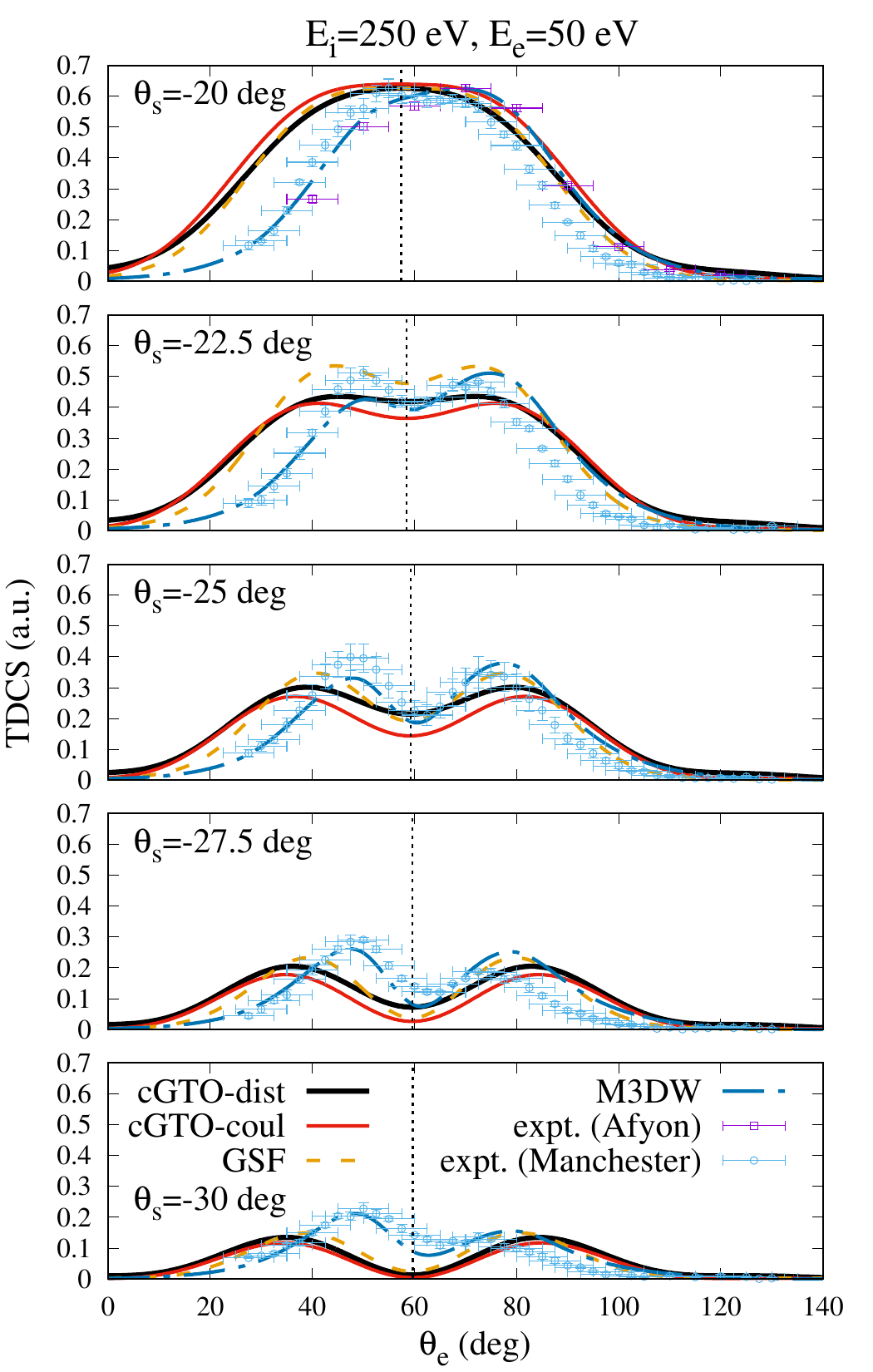}
	\caption{
\ac{TDCS} for ionization by electronic impact of the outer (1t$_2$) valence orbital of CH$_4$, as a function of the ejection angle $\theta_{e}$, for different values of the scattering angle (from top to bottom) $\theta_s = -20^{\circ}$, $-22.5^{\circ}$, $-25^{\circ}$, $-27.5^{\circ}$ and $-30^{\circ}$
with kinematical parameters of Ali {\textit{et al}}~\cite{Ali_2019}, $E_i = 250$~eV, $E_e = 50$~eV.
The present cGTO calculations with the ejected electron described by a Coulomb wave (red line) and a distorted wave (black thick line), are compared with the theoretical (M3DW - dashed blue  and GSF - dashed orange) curves and with the experimental points published in \cite{Ali_2019}. For $\theta_s = -20^{\circ}$, the experimental points of \cite{icsik_2016} are also shown. The vertical dotted lines indicate the momentum transfer direction. 	
		The experimental data, as well as the M3DW and the GSF results, have been normalized to the cGTO curve obtained with distorted potential at the binary peak maximum for $\theta_s = -20^{\circ}$.
	}
	\label{fig:CH4_ali_50eV}
\end{figure}

\begin{figure}[]
\includegraphics[width=\linewidth]{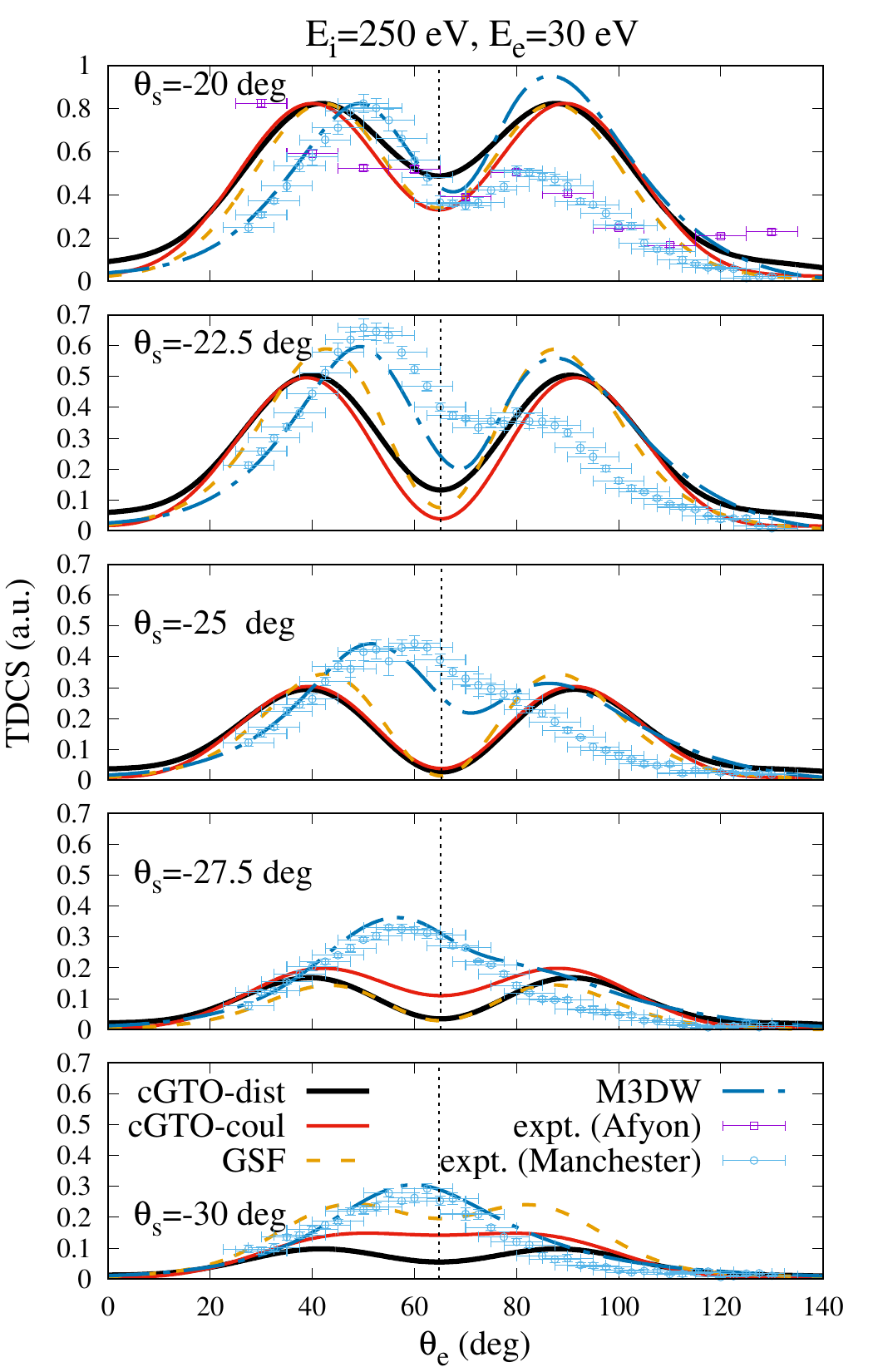}
	\caption{
	Same as Fig. \ref{fig:CH4_ali_50eV} but for $E_e = 30$ eV. 		The experimental data, as well as the M3DW and the GSF results, have been normalized to the cGTO curve obtained with distorted potential at the binary peak maximum for $\theta_s = -20^{\circ}$.
	}
	\label{fig:CH4_ali_30eV}
\end{figure}

As a second test, we have explored the kinematical parameters presented in the more recent experiments of Ali \textit{et al} \cite{Ali_2019}, {\textit{i.e.}} an incident electron energy of 250~eV, an ejected electron with either energy 50~eV or 30~eV, and several fixed scattering angles of the fast outgoing electron ($\theta_s$ between -20$^{\circ}$ and -30$^{\circ}$ with $2.5^{\circ}$ steps). The low energy electrons are detected over the angular range $\theta_e \in [27.5 ^{\circ}, 130 ^{\circ}]$, with a focus on the binary region.
%
%
The cGTO TDCS are presented in Fig.~\ref{fig:CH4_ali_50eV} for the target orbital $1t_2$ with an ejected electron energy of 50 eV. Fig. \ref{fig:CH4_ali_50eV}  also collects the experimental data and some of the  theoretical results of \cite{Ali_2019}.
In this experiment, the data associated with different scattering angles were internormalized so that only one point, the maximum of the $\theta_s = - 20^{\circ}$ curve, was used to normalize the experimental data to the distorted wave calculation for all panels of Fig. \ref{fig:CH4_ali_50eV}.
The earlier measured data of \cite{icsik_2016} at $\theta_s = -20 ^{\circ}$ are also reported.
%
The distinctive feature of the TDCS from $1t_2$ state is the splitting of its binary peak into a double peak appearing with increasing scattering angle, due to the p-type character of this orbital, as previously predicted in \cite{granados_2017} and observed in \cite{Ali_2019}.
Again, the present cGTO results compare quite well with the experimental results. The double peak structure is well reproduced. As expected, the angular shift of the binary peak from the direction of the momentum transfer direction is not recovered due to the first Born approximation.
%
Comparison within theoretical results remains very satisfactory, the cGTO calculation being close to the results of
earlier calculations using Generalized Sturmian Functions (GSF) \cite{Ali_2019,granados_phd}. Both are symmetric with respect to the momentum transfer, and differ from the results obtained with the
molecular three-body distorted wave (M3DW) approach  \cite{Ali_2019,madison_2010}. This is to be expected since the latter uses for the initial bound state a Dyson molecular orbital for the active electron, and for the final state two distorted waves multiplied by an electron-electron distortion factor; moreover, contrary to our case, exchange is included.
The discrepancies between the theoretical methods (GSF and M3DW), and between theoretical and experimental points, have been analyzed in \cite{Ali_2019}.
The present results confirm this global picture with a similar agreement as the orders of magnitude are concerned.
Indeed, since the TDCS have been internormalized,
Fig.~\ref{fig:CH4_ali_50eV} provides
not only a visual comparison of the shapes but also an insight of the TDCS relative magnitudes at different scattering angles.

%
Similar conclusions can be drawn from the analysis of Fig. \ref{fig:CH4_ali_30eV} where the TDCS is shown for an ejected electron energy of 30 eV, all other parameters remaining unchanged. With this lower ejection energy, the agreement between theory and experiment is better at low scattering angles. Discrepancies appear at larger scattering angles when the double peak fades in the experimental data but remains in the present cGTO (Coulomb or distorted) and GSF theoretical curves.

TDCS calculation using cGTO-expanded wave functions and analytical integrals thus succeed in reproducing the main features of experimental $(e,2e)$ TDCS for methane in coplanar asymmetric geometry and with different sets of kinematic parameters. These results validate the present Gaussian methodology as a reliable tool for the theoretical description of collision processes involving continuum states.



\section{Summary \label{sec_summary}}

We have studied theoretically the ionization  of the inner ($2a_1$) and outer ($1t_2$) valence orbitals
 of the CH$_4$ molecule, by both photon and electron impact. Inspired by the available experimental data, we have considered here ejected electrons with energy up to about 2.7 a.u..
 For the photoionization, within the dipolar approximation, we have looked at the energy dependence of the cross section and the asymmetry parameter related to the angular distribution of the ejected electron.
 For the electron impact ionization, within the first Born approximation, we have focused on the angular distributions of the triple differential cross sections related to $(e,2e)$ experiments in coplanar asymmetric geometries, in which the incident electron is scattered with an energy much larger than the ejected electron.
 For both processes, we have worked within a one-center approach, and with bound molecular target states described by Slater type orbitals provided in the literature.
 The ejected electron is described by a continuum state of a model molecular central potential; the corresponding radial function is subsequently represented by a finite sum of complex Gaussian type orbitals, \textit{i.e.}, GTO with complex-valued exponents.
 With these ingredients, we have provided the formulation to calculate  all ionization matrix elements analytically.
The present manuscript  has given solid evidence that the calculation scheme works very well.

 In spite of the approximations, the calculated ionization observables for CH$_4$ are of fair quality.
 For photoionization, when compared to other theoretical results and experimental data, an overall acceptable agreement is found in the velocity gauge, especially for the outer $1t_2$ orbital.
 For electron impact ionization, and for different sets of kinematical and geometrical parameters, a reasonable agreement is found with TDCS obtained by other theories using similar approximations and also with measured angular distributions.

In our opinion, more importantly than the calculated cross sections it is the approach itself which opens new perspectives.
The originality stands in the use of cGTO to represent the continuum radial function. In the present study such representation is demonstrated to be sufficiently accurate for energies up to 2.7 a.u. and spanning a radial domain of 30 a.u.. It is used here in combination with one-center molecular bound states described by a sum of Slater type orbitals, leading to ionization matrix elements in closed form. The analytical formulation is even simpler if monocentric Gaussian type orbitals are used.
With practically no effort, it can be readily adapted for both ionization processes, the radial integrals reducing to well known Gaussian integrals (we refer to \cite{Ammar_2023_integrals} for the formulas and a numerical investigation).

The proposed approach allows one to envisage treating molecular processes, for example with large molecules as long as the wave function is practically negligible beyond 30 a.u.
(as was the case for a small molecule such as CH$_4$).
Moreover, the analytical character of the matrix elements is maintained also in a multicentric GTO description of the target.
In that case, the formulation, whether in radial or Cartesian coordinates, makes use of the Gaussian mathematical properties extended to the complex plane \cite{Ammar_2021_Multicenter}.
As a consequence our Gaussian approach should allow us to consider studying molecular processes with molecules whose orbitals are issued for example by the Gaussian package \cite{g16}. Finally, it is our intention to further develop the method by considering multicentric continuum states, whose cGTO representation would lead to an all-Gaussian approach.





\section*{References}


%

\end{document}